\documentclass[preprint,review,12pt]{elsarticle}

\usepackage{tikz}
\usepackage{textcomp}
\usepackage{hyperref}
\usepackage{float}
\usepackage{multirow}
\usepackage{subfigure}

\usepackage{amssymb}

\usepackage{algorithm}
\usepackage{algorithmic}

\usepackage{latexsym}

\usepackage{hyperref}

\usepackage{comment}
\usepackage{caption}
\usepackage[T1]{fontenc}
\usepackage[utf8]{inputenc}
\usepackage[english]{babel}
\newcommand\copyrighttext{%
  \footnotesize \textcopyright 2021 IEEE. Personal use of this material is permitted.
  Permission from IEEE must be obtained for all other uses, in any current or future 
  media, including reprinting/republishing this material for advertising or promotional 
  purposes, creating new collective works, for resale or redistribution to servers or 
  lists, or reuse of any copyrighted component of this work in other works. 
  DOI: \href{<https://ieeexplore.ieee.org/document/9579004>}{10.1109/TIFS.2021.3121201}}
\newcommand\copyrightnotice{%
\begin{tikzpicture}[remember picture,overlay]
\node[anchor=south,yshift=10pt] at (current page.south) {\fbox{\parbox{\dimexpr\textwidth-\fboxsep-\fboxrule\relax}{\copyrighttext}}};
\end{tikzpicture}%
}

\journal{Pattern Recognition}

\begin{document}

\begin{frontmatter}

\title{Fingerprint recognition with embedded software-based presentation attacks detection: are we ready?}

\author{Marco Micheletto, Gian Luca Marcialis, Giulia Orrù, and Fabio Roli}
\address{Department of Electrical and Electronic Engineering\\University of Cagliari\\Piazza d'Armi - I-09123 Cagliari (Italia)}
\ead{\{marco.micheletto, marcialis, giulia.orru, roli\}@unica.it}



\begin{abstract}
\copyrightnotice

The diffusion of fingerprint verification systems for security applications makes it urgent to investigate the embedding of software-based presentation attack detection algorithms (PAD) into such systems. Companies and institutions need to know whether such integration would make the system more “secure” and whether the technology available is ready, and, if so, at what operational working conditions. Despite significant improvements, especially by adopting deep learning approaches to fingerprint PAD, current research did not state much about their effectiveness when embedded in fingerprint verification systems. 
We believe that the lack of works is explained by the lack of instruments to investigate the problem, that is, modeling the cause-effect relationships when two non-zero error-free systems work together. Accordingly, this paper explores the fusion of PAD into verification systems by proposing a novel investigation instrument: a performance simulator based on the probabilistic modeling of the relationships among the Receiver Operating Characteristics (ROC) of the two individual systems when PAD and verification stages are implemented sequentially. As a matter of fact, this is the most straightforward, flexible, and widespread approach. 
We carry out simulations on the PAD algorithms' ROCs submitted to the most recent editions of LivDet (2017-2019), the state-of-the-art NIST Bozorth3, and the top-level Veryfinger 12 matchers. Reported experiments explore significant scenarios to get the conditions under which fingerprint matching with embedded PAD can improve, rather than degrade, the overall personal verification performance.
\end{abstract}




\end{frontmatter}


\section{Introduction}
\label{sec:intro}

Software-based detection of fingerprint presentation attacks (\cite{sousedik2013survey,marasco2014survey,marcel2014handbook})
is also called fingerprint liveness detection, or fingerprint anti-spoofing\footnote{In general, a spoofing attack is only a possible kind of presentation attack, but, in the case of fingerprints, these terms coincide.}. It has been boosted in the last ten years thanks to the availability of datasets that led to extensive deep networks training, whose adoption is considered the most innovative and promising approach at the state-of-the-art.

\begin{figure}[tb]
   \centering
   \includegraphics[width=\linewidth]{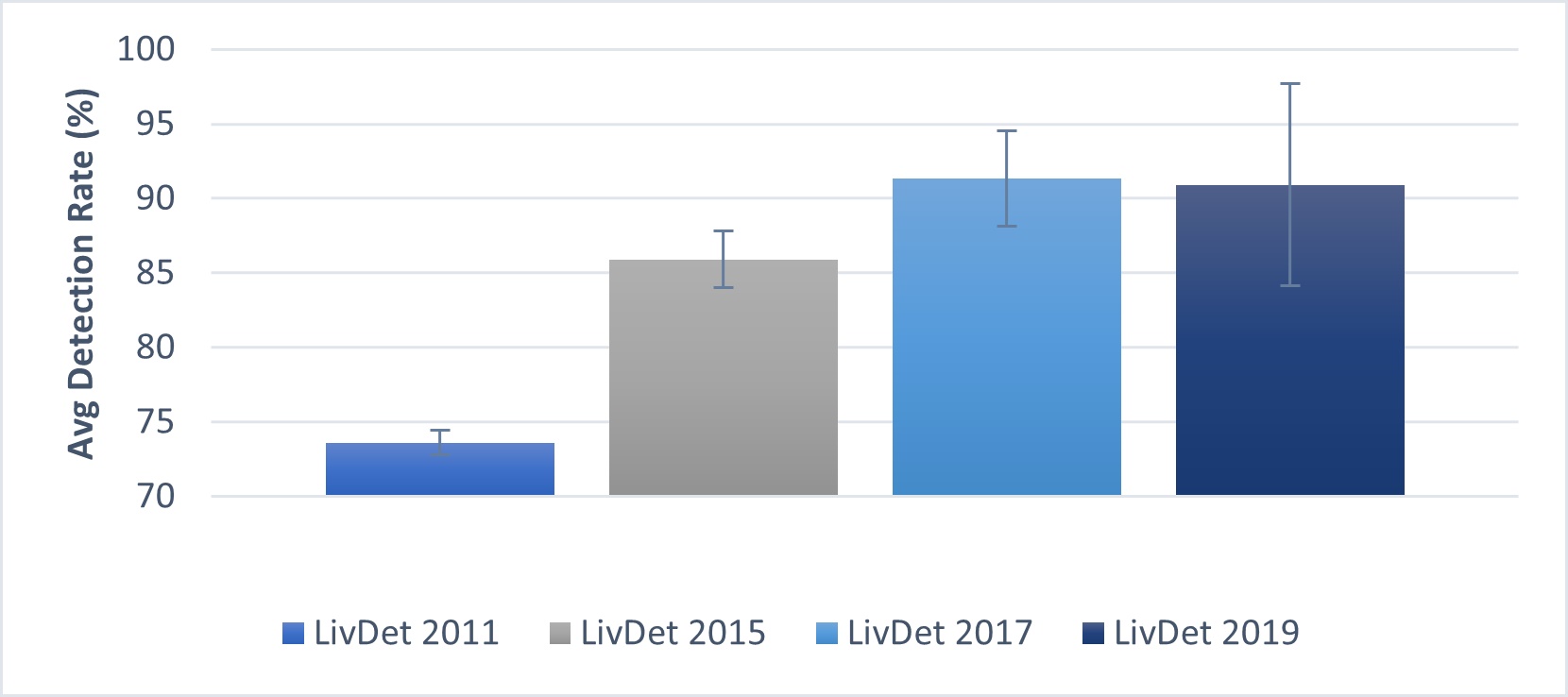}
   \caption{Percentage accuracy and related standard deviation over the datasets and participants to 2011-2019 editions of LivDet \cite{mura2015livdet,ghiani2017livdetReview}.}
\label{fig:livdetOverYears}
\vspace{-10pt}
\end{figure}

Many publications focused on showing the performance improvement and the number of datasets for this aim enormously increased. Among other initiatives, the International Fingerprint Liveness Detection competition, also known as LivDet, is a biennial appointment for academies and companies to make the point on PAD algorithms. The authors of this paper are co-founders and co-chairs of LivDet 2009-2015 \cite{ghiani2017livdetReview}, and chairs of LivDet 2017-2021 (the last edition is ongoing)\footnote{http://livdet.diee.unica.it}.
Fig. \ref{fig:livdetOverYears} summarizes the main achievements of LivDet from 2011 to 2019 \cite{ghiani2017livdetReview,mura2015livdet,orru2019livdet}. It is easy to see that the overall average accuracy of PAD algorithms in the last two editions was around 90\%. This means that an average PAD missed 10\% of presentation attacks as well as \textit{bona fide} presentations, according to the ISO terminology \cite{busch2019HOA}. The acceptability of this error rate depends on the application. Let us consider the common application scenario of embedding a PAD into a fingerprint verification system, a critical task for digital banking, online shopping, classified documents protection. LivDet2019 also evaluated this scenario. The best three fingerprint verification algorithms with embedded PAD submitted to the competition reported an average equal error rate of $4.85\%\pm 3.69\%$ \cite{orru2019livdet}. This performance is much worse than that usually claimed by the best off-the-shelf fingerprint matchers when neglecting presentation attacks. For example, the reliability tests of VeryFinger 12 reported an Equal Error Rate of $0.882\%$ in the worst case\footnote{\url{https://www.neurotechnology.com/verifinger-algorithm-tests.html}}. Unfortunately, no other recent works explored PAD and matching systems' embedding; thus, reported values could appear limited. However, we believe that this is sufficient to consider the problem relevant to the future application of large-scale fingerprint recognition systems. As a matter of fact, the state of current software-based PAD systems \cite{kolberg2020} may suggest vendors consider software-based PAD as a marginal feature though necessary for guaranteeing the product placement into the market. Eventually, hardware-based PAD would be preferred \cite{Goicoechea2019}.

Actually, the literature reports a few of investigations on the PAD-matcher embedding problem \cite{abhyankar2009integrating,marasco2012combining,rattani2013bayes,marcel2014handbook,marcel2013action}. Parallel and sequential combinations of PAD and matcher are just present in Refs. \cite{abhyankar2009integrating,marasco2011incorporating,marasco2012combining,rattani2012fusion,rattani2013bayes,wong2014attack}, not limited to fingerprints. The 2nd edition of the \textit{Handbook of Biometric Anti-spoofing} reported an evaluation methodology for integrated systems (Chp. 12, \cite{marcel2019HOA}). All the above cited works acknowledged the existence of such performance degradation \cite{abhyankar2009integrating,marcel2013action}. However, it is not yet clear \textit{how much} is this degradation. Is the integrated system still \textit{effective} for security applications? 
Up to now, the only thing to do is collecting data, replicating algorithms, performing experiments, and evaluating the embedding feasibility time-to-time. In no other way can we estimate the error rates and the system's effectiveness according to the adopted sensor, the spoofing materials, the probability of a presentation attack \cite{marcel2019HOA}. While the sensor characteristic and material adopted are the main variables to evaluate a PAD system, the literature never considered the impact of the spoofing attack probability in all aspects. For example, high-security applications and consumer ones (sensors embedded into smartphones) have different goals. In the second ones, vendors could hypothesize that presentation attacks are much less likely or not relevant. From the scientific viewpoint, the reason is the same for which a few works are focused on embedding PADs into fingerprint recognition systems: there is no other way to evaluate current PAD performance when embedded into verification systems than that mentioned above.

Things could change if a tool modeling the presentation attack probability were available. The tool would allow evaluating, for example, the impact of latex-made PAIs\footnote{Presentation Attack Instrument or, simply, spoof or fake fingerprint.} and the sensor technology on the whole system's ROC. We would be able to decide for which operational points and conditions the given embedding is worthy to be implemented or not. The designer would have a handy tool to assess the performance of high-security and consumer applications.

This is what we propose in the paper: a simulator based on the probabilistic modeling of relationships among variables at hand in the case of the \textit{sequential} fusion of presentation attacks detector and matcher. Sequential fusion is only a possible choice, but it is the simplest and most widespread one. 
It is also very flexible because, for example, when a better PAD is available, we can substitute it into the system without caring about the matcher. 
Moreover, this type of integration allows us to efficiently model the link between performances and probabilistic relationships under appropriate working hypotheses.
Other, more complex ways to embed a PAD into a fingerprint verification system are also more challenging to handle \cite{marasco2011incorporating,marasco2012combining,rattani2012fusion,rattani2013bayes,wong2014attack}, and it has been not yet shown, neither theoretically nor experimentally, the superiority of such approaches to this one.

Our simulator takes as input the ROC curves of the fingerprint matcher and the PAD. The output is the whole acceptance rate in its three essential components: the genuine users one (GAR), the zero-effort attacks one (FMR), and the presentation attacks one (IAPMR)\cite{abhyankar2009integrating}. Two parameters are added: the probability of a presentation attack and the PAD's operational point. 
Using this simulator does not require implementing or replicate any PAD algorithm or matching system. This is also what a designer would prefer to do: take the vendors' individual ROCs and explore the performance achievable according to some expected scenarios. To show the pros and cons of our simulator, we considered the PAD algorithms of LivDet 2017-2019 as good representatives of the state of the art, due to the high PA detection rates individually attained. Moreover, they adopted convolutional neural networks, acknowledged mainly as the PAD systems of ``novel generation''. We computed their ROC curves over the LivDet 2017-2019 datasets and investigated limits and potentials of their embedding with the well-known NIST Bozorth3, that is, the main benchmarking matcher publicly available\footnote{https://www.nist.gov/services-resources/software/nist-biometric-image-software-nbis} and the Veryfinger 12, namely, the top-level matcher nowadays off-the-shelf\footnote{https://www.neurotechnology.com/verifinger.html}. These matchers may represent a medium-level security scenario and a high-level security scenario, respectively. 
After verifying the reliability of our simulator in terms of the difference between expected and real acceptance rate values, the impact of sensors and materials used for making spoofs was widely investigated and explained. The simulator also allowed to derive the main guidelines to follow for deciding whether a particular PAD can be embedded in the matching process.

The paper is organized as follows. Section 2 introduces the terminology and reviews the current literature in order to motivate our contribution. Section 3 describes the theoretical model. Section 4 reports the simulations. Conclusions are done in Section 5.

\section{Current literature and our work's motivation}
\label{sec:SOTA}
First of all, we give the terminology recently introduced in order to take into account the presence of both zero-effort (impostors submitting their own fingerprint) and presentation attacks (impostors submitting a fake fingerprint of the targeted user). In accordance with the metrics defined within the recent ISO / IEC 30107-3 standard for the PAD evaluation \cite{busch2019HOA}, we refer to error rates in terms of Attack Presentation Classification Error Rate (APCER), \textit{Bona fide} Presentation Classification Error Rate (BPCER).
The usual False Match Rate (FMR) and Genuine Acceptance Rate (GAR) are coupled with the term Impostor Attack Presentation Match Rate (IAPMR) \cite{abhyankar2009integrating}, which is the False Match Rate under presentation attacks, being FMR referred to zero-effort attacks only. 


%

Therefore, which are the main achievements in literature? The most relevant points are briefly summarised.
Proposed integration approaches are based on sequential fusion \cite{abhyankar2009integrating} or try modeling liveness and match scores by probabilistic relationships between measurements and events in the form of a Bayesian network \cite{marasco2012combining,rattani2012fusion,rattani2013bayes,wong2014attack}. The only work adopting sequential fusion is Ref. \cite{abhyankar2009integrating}, where tests are carried out with a few samples to allow any significant conclusion or insight about the general performance of the system. Nevertheless, this is a pioneering work, which was not followed by other analyses.

None of the above papers helps in explaining the relationships among FMR, GAR and IAPMR of the integrated system. It is even impossible to estimate a priori the amount of the GAR decrease by those models, that is, by knowing the error rates of the individual systems. Concerning this issue, Ref. \cite{marcel2014handbook} (p. 471) showed that the weighted sum of FMR and IAPMR, named $FAR_{\omega}$ is reduced when the matcher is integrated with a PAD system. However, GAR is still lower than that achieved by getting rid of presentation attacks. Again, when and why this may happen is still unexplained.
A first experimental and statistically relevant evidence of PAs on a fingerprint matching system can be found in \cite{biggio2017analysis}, where it can be noticed that $IAPMR >>FMR$ for the same value of acceptance threshold (p. 568). In \cite{biggio2017analysis}, this is noticeable although a robust multi-modal rule is used to reduce this effect. Moreover, the plots of p. 568 show that GAR can lower up to 40\% and more when the fingerprint sensor is subjected to presentation attacks.
\\
Finally, according to LivDet results, the performance of current PADs (Fig. \ref{fig:livdetOverYears}) suggests approaches as hardware-based liveness detection or tricks like enrolling multiple fingers \cite{crossmatch2014}, especially by considering large scale applications. This cannot be neglected even though the use of deep networks led to better performances than those of hand-crafted features-based PADs \cite{nogueira2016TIFS}. In particular, the performance of LivDet 2015 is mostly related to the use of hand-crafted features \cite{mura2015livdet}, while the LivDet 2019's is mainly related to the use of convolutional neural networks \cite{ghiani2017livdetReview}, and these are used in almost all algorithms submitted to the 2019 edition \cite{orru2019livdet}. 

The main problem of integration is that many things like the relationships between GAR, IAPMR, and FMR of the fingerprint matcher, when combined with the PAD’s ROC, cannot be obtained without collecting data, replicating algorithms, and performing experiments. Therefore, representing the best case or the worst-case scenario, for example, due to a large number of presentation attacks over time or specific attacks adopting some particularly insidious material for the sensor is out of the current possibilities.

This motivates the need for a specific module, namely, a simulator of possible scenarios, able to present the overall system performance under integration without the implication of the practical difficulties above. Therefore, this paper proposes a simulator explicitly designed to model the sequential fusion of PAD and verification systems. 
Thanks to this instrument, the designer can understand whether sequential fusion works depending on the kind of attacks (materials), and the probability of being attacked. Moreover, she/he can state ``where we are'' according to the current technology on which matchers and PADs are based.

\section{Modeling and simulation of fingeprint recognition system with embedded PAD}

\begin{figure}[tb]
   \centering
   \includegraphics[width=\linewidth]{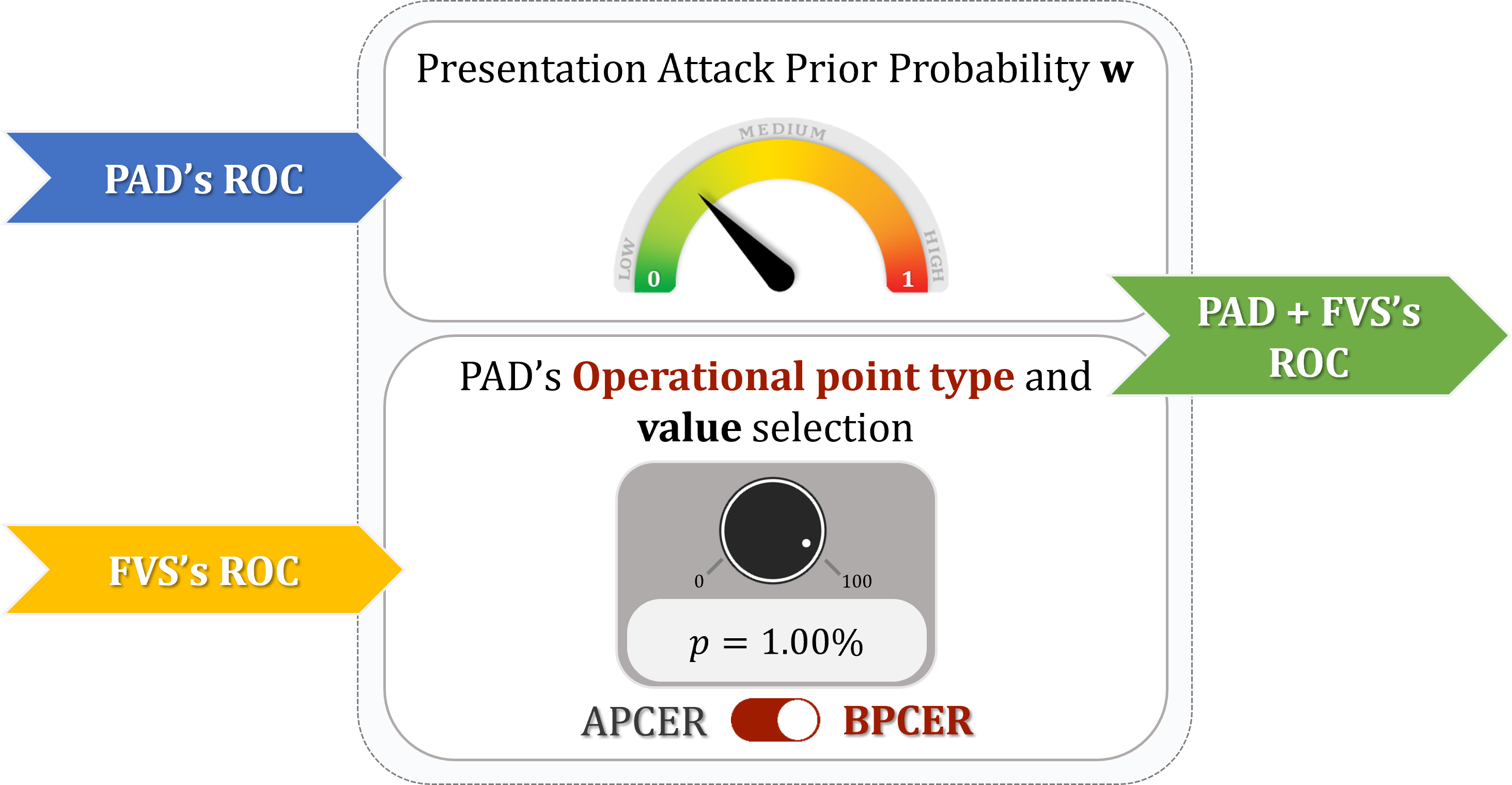}
   \caption{Scheme of the proposed simulator. It takes as input the ROC’s of the individual fingerprint verification system (FVS) and PAD modules and provides the global ROC of fingerprint recognition system with embedded PAD.}
\label{fig:simulator}
\vspace{-10pt}
\end{figure}

According to what was previously stated, this work aims to propose a simulator able to deal with returning the performance expectation of the sequential fusion of fingerprint matching and PAD. 
Fig. \ref{fig:simulator} shows the simulator's high-level view. It takes as input the individual ROCs of the matcher and the PAD and provides the ROC of their fusion. The simulator allows to investigate the performance according to two parameters: the prior probability of being attacked by spoofs, namely, $w$ in this paper (see also the term $\omega$ in Ref. \cite{marcel2019HOA}), and the specific operational point chosen for the PAD, set by $BPCER=p\%$ or $APCER=p\%$. The system can also return the performance by varying $w$ to set when it is convenient to turn on/off the PAD module.

In the following sections, after introducing the needed terminology, we show how this simulator can be implemented by appropriate modeling of the individual ROCs and the role of $w$.

\subsection{Problem modeling}
\label{subsec:term}

First of all, let $G$ be the boolean event ``the input user is authorized''. Therefore, $\bar{G}$ indicates the opposite event. Obviously $G \cup \bar{G}=\Omega$ where $\Omega$ is the \textit{Certain Event}.
Secondly, let $L$ be the boolean event ``the input image is alive/authentic'', that is, the input image is from the alive fingerprint of the user; $\bar{L}$ indicates that the input image is from a fake/spoof fingerprint. Even in this case, $L \cup \bar{L}=\Omega$. We also indicate with $P(G)$ and $P(L)$ the corresponding probabilities, so that $P(\bar{G})=1-P(G)$ and $P(\bar{L})=1-P(L)$. 

According to this notation, we have the possible joint events:

\begin{itemize}
\item $\{L, G\}$: the input image is alive, and the user is authorized (genuine user trial);
\item $\{ L, \bar{G}\}$: the input image is alive, and the user is unauthorized (zero-effort attack);
\item $\{\bar{L}, \bar{G}\}$: the input image is spoof and the user is unauthorized (presentation or spoofing attack);
\item $\{\bar{L}, G\}$: this event is technically \textit{impossible} because an authorized user is never supposed to use his own replica to gain access to the system\footnote{The event ``a user in the white-list tries to get the access as another authorized user'' falls in the case $\{ L, \bar{G} \}$}.  
\end{itemize}

\begin{figure}[tb]
   \centering
   \includegraphics[width=.5\textwidth]{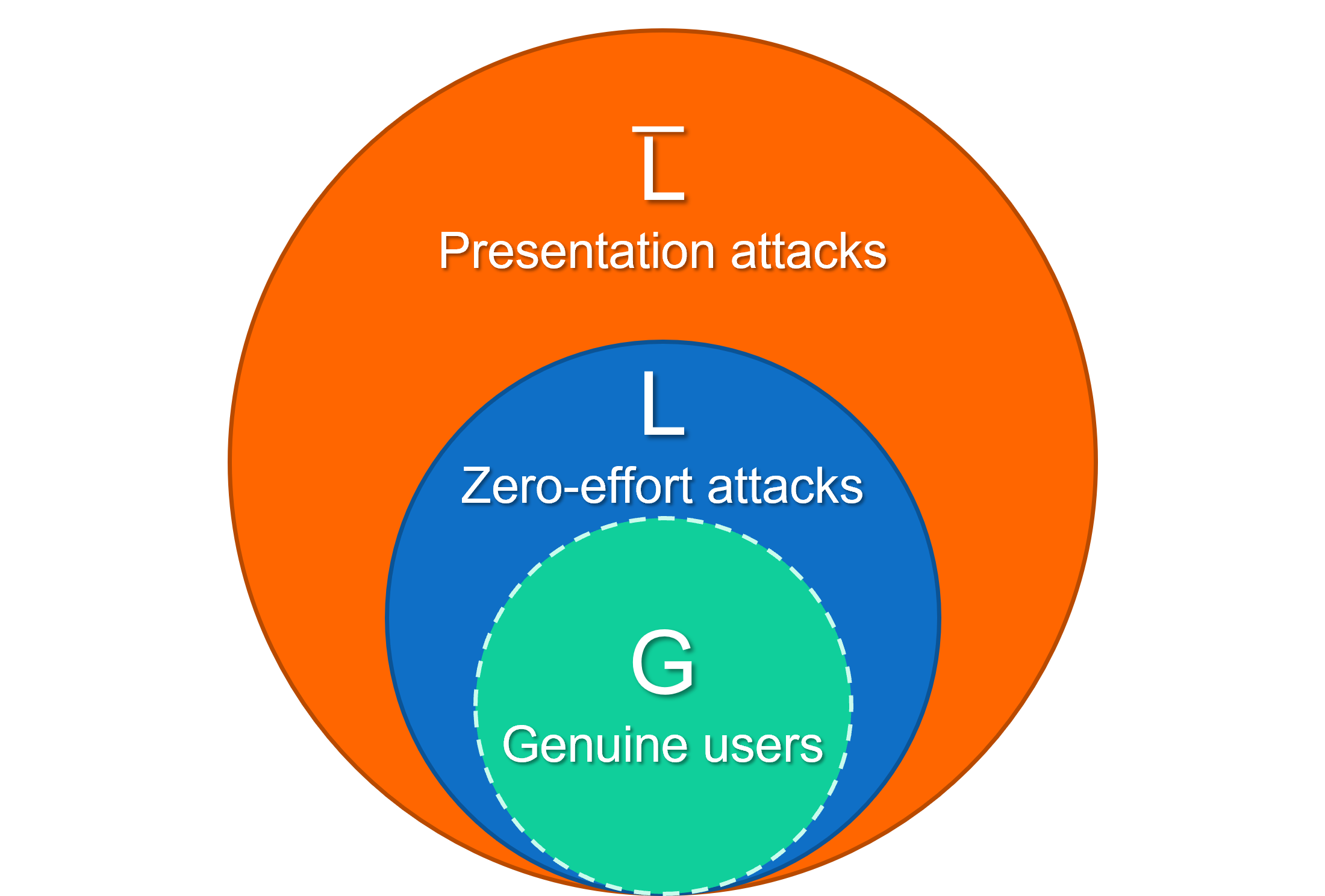}
   \caption{Relationship between G and L events.}
\label{fig:venns}
\vspace{-10pt}
\end{figure}

Fig. \ref{fig:venns} summarizes the relationships between $G$ and $L$ by the Venn's diagram. It points out that $G\subseteq L$: $L$ not only includes all genuine users, but also impostors which tries zero-effort attacks.

In order to model the acceptance rate of a single matcher and a liveness detector appropriately, we introduce two events driven by the output of the matching and the liveness detection phases.
As written in the introduction, the access is granted to a certain user when the matching score $s_M$ between the input image and the user's claimed identity template(s) is over a given acceptance threshold $s_M^*$. For the sake of brevity, we will define the following boolean event:
\begin{eqnarray}
\label{eq:acceptancerate}
M=s_M>s_M^*
\end{eqnarray}

Accordingly, $P(M)$ is the acceptance probability of a generic input sample, that is, the \textbf{acceptance rate}. In our modeling, this probability is independent of the specific user. Therefore, no user-specific approach is taken into account here without loss of generalization. We can also write this expression as a function of $L$ and $M$ events:

\small
\begin{eqnarray}
\label{eq:arfirst}
P(M)=P(M|G,L)\cdot P(G,L)+ 
P(M|\bar{G},L)\cdot P(\bar{G},L)+ \nonumber\\
P(M|G,\bar{L})\cdot P(G,\bar{L})+
P(M|\bar{G},\bar{L})\cdot P(\bar{G},\bar{L})
\end{eqnarray}
\normalsize

Similarly, the liveness detector gives the classification of a certain input sample as alive or fake when the liveness score $s_F$, obtained by the analysis of the feature set extracted from the input image, is over a certain liveness threshold $s_F^*$. Therefore, we define the following event:
\begin{eqnarray}
\label{eq:acceptancerate}
F=s_F>s_F^*
\end{eqnarray}
$P(F)$ is the general probability of classifying a generic pattern as alive.

On the basis of the definition above we may represent the acceptance rate of each access trial for the individual matcher:

\small
\begin{eqnarray}
\label{eq:acceptancerate}
P(M|G,L)=GAR(M)\\
P(M|\bar{G},L)=FMR(M)\\
P(M|\bar{G},\bar{L})=IAPMR(M).
\end{eqnarray}
\normalsize

Where GAR, FMR, IAPMR are the so-called Genuine Acceptance Rate, False Match Rate, and Impostor Attack Presentation Match Rate, respectively.

At the same time, we may represent the \textit{bona fide} and presentation attack classification error rates of a liveness detector:
\small
\begin{eqnarray}
\label{eq:livenessrate}
P(F|L)=1-BPCER(F)\\
P(F|\bar{L})=APCER(F).
\end{eqnarray}
\normalsize

\subsection{The proposed simulator}
\label{subsec:model}

\begin{figure}[tb]
   \centering
   \includegraphics[width=.5\textwidth]{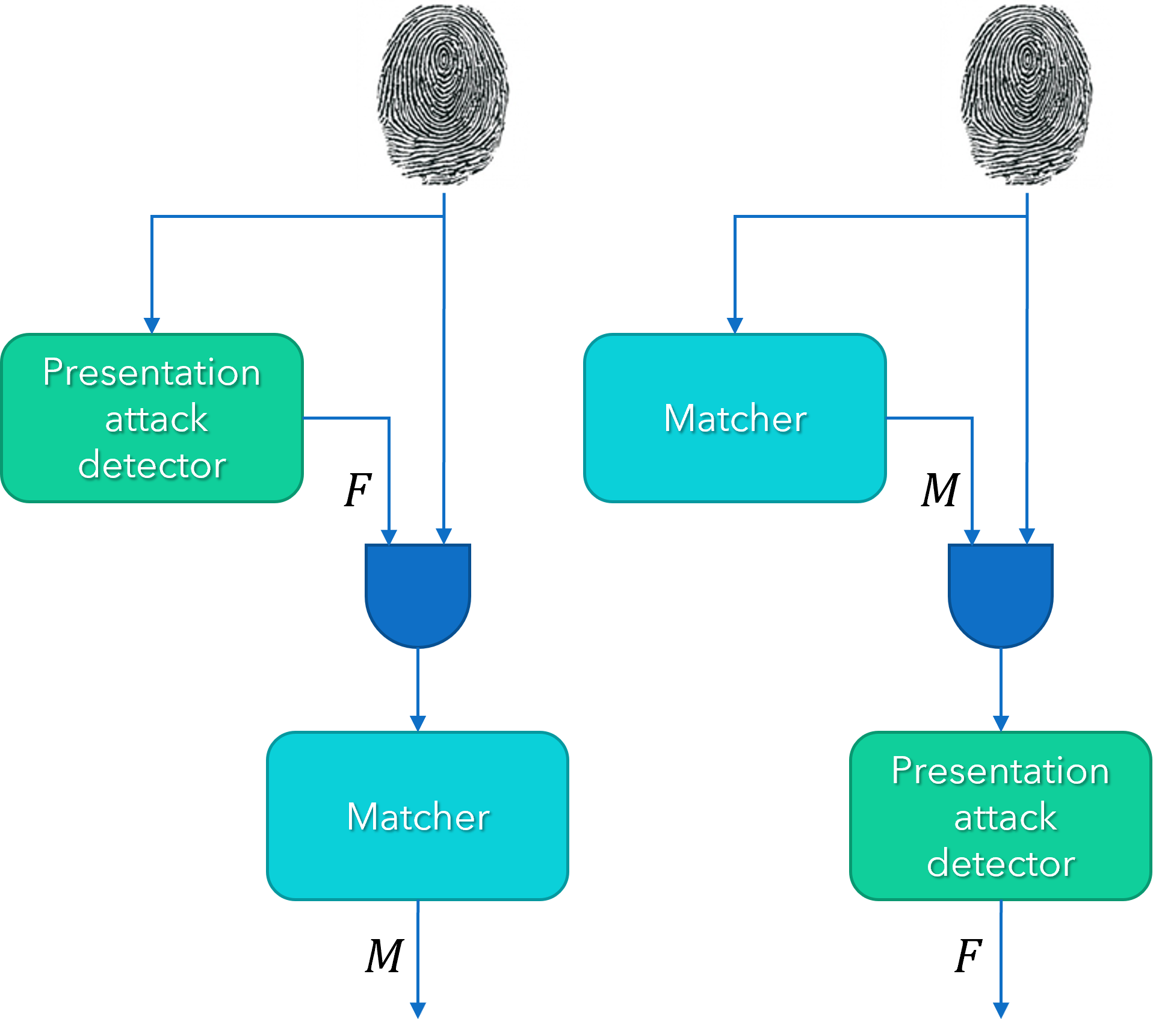}
   \caption{Serial combination of PAD and matcher.}
\label{fig:serialsystem}
\vspace{-10pt}
\end{figure}

A sequential system is depicted in Fig. \ref{fig:serialsystem}. We may have two possibilities: the liveness detection module precedes or succeeds the matcher. Therefore, the probability of acceptance given a particular state of nature $\{ L, G\}$ is constrained to the verification of events $F$ and $M$. In the following, we model the expression of the acceptance rate by using the terminology previously introduced. We avoid specifying the actual value of truth associated with $L$ and $G$ since equivalent expressions can be obtained for each configuration of these random variables.\newline 

\subsubsection{Matching and Liveness}
\label{subsec:matchliveness}

In this architecture, the matching check preceeds the liveness check. Accordingly, the probability of acceptance given a specific configuration of $\{ L, G \}$, is:
\begin{equation}
\label{eq:LMg}
P(M,F|L,G)=P(M|L,G)\cdot P(F|L,G,M)
\end{equation}

The following hypotheses can simplify this complex expression.

\textbf{Hypothesis 1}: \textit{F is independent of G given M and L}:
\small
\begin{equation}
\label{eq:LMind}
P(F|L,G,M)=P(F|L,M)
\end{equation}
\normalsize

Motivation: being $G$ a subset of $L$, as depicted in Fig \ref{fig:venns}, and remembering that $F\neq L$, his probability $P(F)$ depends only on the liveness threshold $s_F^*$ and the liveness score. Due to presentation attacks existence, if a match is found ($M=1$), the pattern can be in the $L$ or $\bar{L}$ sides, regardless of the truth level of $G$. Therefore we can state that no information about the \textit{liveness probability} is given by the evidence that the sample belongs to an authorized user. \newline

\textbf{Hypothesis 2}: \textit{F is independent of M, given L}:

\small
\begin{equation}
\label{eq:LMhyp}
P(F|L,M)=P(F|L)
\end{equation}
\normalsize
Motivation: it is unreasonable to suppose a probabilistic relationship between the liveness score and the fact that the sample may match or not. A certain pattern may be classified as alive independently of the fact that it matched with the claimed identity's template. This agrees with the definition of liveness detector module, which must detect the authenticity of the fingerprint independently of the user population.

Therefore:
\small
\begin{equation}
\label{eq:LMs}
P(M,F|L,G)=P(M|L,G)\cdot P(F|L)
\end{equation}
\normalsize

To sum up, we may express the acceptance rate of this architecture by the performance and security parameters of the individual matching and liveness modules:

\small
\begin{eqnarray}
\label{eq:AR_LM}
GAR_{Match\rightarrow Live}=GAR(M)\cdot (1-BPCER(F)) \nonumber\\
FMR_{Match\rightarrow Live}=FMR(M)\cdot (1-BPCER(F))\\
IAPMR_{Match\rightarrow Live}=IAPMR(M)\cdot APCER(F)\nonumber
\end{eqnarray}
\normalsize

The expressions above point out that, whilst FMR and IAPMR decrease with respect to those of the individual macher, GAR decreases too. This theoretically confirms what was reported experimentally \cite{abhyankar2009integrating}. Thanks to this modeling, we can point out that:
\begin{enumerate}
    \item GAR always decreases in the integrated system if we keep the same operational point for the matcher.
    \item The amount is inversely proportional to the BPCER of the presentation attacks detector.
\end{enumerate}
Therefore, by this instrument, it is also possible to evaluate how much GAR decrease is expected.\newline

\subsubsection{Liveness and Matching}
\label{subsec:livenessmatch}

In this architecture, the liveness check is done before the matching check: 

\small
\begin{equation}
\label{eq:ML}
P(M,F|L,G)=P(F|L,G) \cdot P(M|L,G,F)
\end{equation}
\normalsize

By recalling that F is independent of G given L (Hypothesis 1 in Section \ref{subsec:matchliveness}):

\small
\begin{equation}
P(M,F|L,G)=P(F|L) \cdot P(M|L,G,F)
\label{eq:ML}
\end{equation}
\normalsize

The main problem is to understand the degree of dependence between $M$ and $F$ events. In other words, what is the probabilistic dependence of obtaining a matching score higher than the acceptance threshold, given that the submitted fingerprint image achieved a liveness score higher than the liveness threshold?

The literature often pointed out that the quality of spoof images is often less than that of the quality of the corresponding alive images. In other words, it is most probable to have a bad fake image than a bad alive image. Based on this claim, several works tried to use quality measurements to disguise between spoof and alive fingerprint images\cite{sousedik2013survey,marasco2014survey,marcel2014handbook,galbally2012quality}.

\begin{figure}[tb]
   \centering
   \subfigure[]{\includegraphics[width=.45\textwidth]{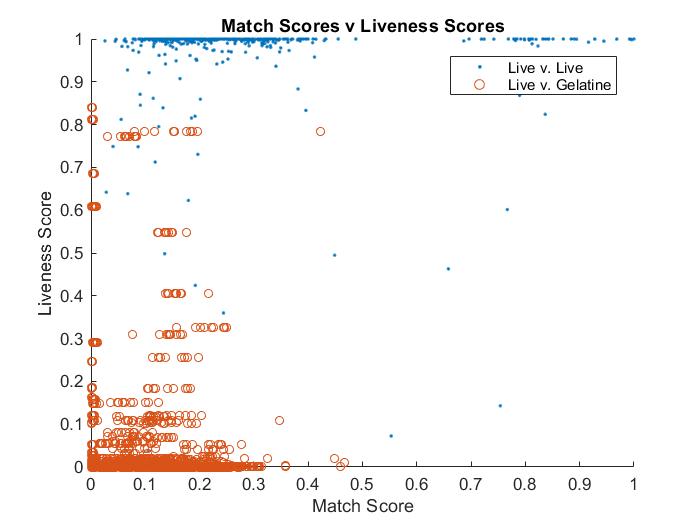}}
   \subfigure[]{\includegraphics[width=.45\textwidth]{GelatineMvL.jpg}}   
   \caption{Quality/liveness scores vs. match scores from the Livdet2013 datasets, when the gelatine is used as cast. It is evident there is not significant correlation among such measurements.These plots are both quoted from  \cite{marasco2018ipta} where more details are available.}
\label{fig:QLvMLivDet2013}
\vspace{-10pt}
\end{figure}

However, a very recent investigation reported in \cite{marasco2018ipta} showed that no explicit statistical dependence can be found between match score and liveness score. This may be easily accepted for zero-effort attacks ($\bar{G}$), while for presentation attacks and genuine trials can be assumed based on findings reported in \cite{marasco2018ipta}. For the sake of example, we report in Fig. \ref{fig:QLvMLivDet2013} the plots of liveness and quality scores vs. the correspondent match scores when a presentation attack is performed using a spoof made up of gelatin, namely, one among the most effective materials. Ref. \cite{marasco2018ipta} reports a statistical analysis on the significance about this correlation absence, as well as investigations on other materials. It is easy to see the possibility of concluding the very low correlation between $M$ and $F$ given the presentation attack. In other words, despite this may be considered counter-intuitive, there is no difference between having an alive finger or a fake one on the sensor surface: the probability of a match ($M=1$) is the same.  

Consequently:
\small
\begin{eqnarray}
\label{eq:ML_novelVersion}
P(M|L,G,F)-P(M|L,G,\bar{F})&=0\\
P(M|L,\bar{G},F)-P(M|L,\bar{G},\bar{F}) &= 0\\
P(M|\bar{L},\bar{G},F)-P(M|\bar{L},\bar{G},\bar{F})&=0
\end{eqnarray}
\normalsize
It follows that:
\small
\begin{eqnarray}
P(M|L,G) = \nonumber \\P(M|L,G,F)\cdot P(F|L,G)+
P(M|L,G,\bar{F})\cdot P(\bar{F}|L,G) =\nonumber \\
P(M|L,G,F)\cdot P(F|L)+ P(M|L,G,F) \cdot P(\bar{F}|L) =\nonumber \\
P(M|L,G,F)
\label{eq:ML1_NV}
\end{eqnarray}
\normalsize

Similarly:
\small
\begin{eqnarray}
P(M|\bar{L},\bar{G}) =\nonumber\\
P(M|\bar{L},\bar{G},F)\cdot P(F|\bar{L},\bar{G})+
P(M|\bar{L},\bar{G},\bar{F})\cdot P(\bar{F}|\bar{L},\bar{G}) = \nonumber\\
P(M|\bar{L},\bar{G},F)\cdot P(F|L)+ 
P(M|\bar{L},\bar{G},F) \cdot P(\bar{F}|L) 
 =\nonumber\\
 P(M|\bar{L},\bar{G},F)
\label{eq:ML11_NV}
\end{eqnarray}
\normalsize
Therefore, the final expressions of GAR, FMR and IAPMR are  the same reported in Eqs. \ref{eq:AR_LM}. In the following, we refer to them as $GAR_{Seq}$, $FMR_{Seq}$, $IAPMR_{Seq}$, respectively.\newline 

\subsubsection{The final model}
\label{subsec:finalModel}

On the basis of  the previous Section, and by recalling Eq. \ref{eq:arfirst}, we obtain the acceptance rate:

\small
\begin{eqnarray}
\label{eq:finalModel}
AR_{Seq}(M,F)=P(M,F) = GAR_{Seq}(M,F) \cdot P(G,L) +\nonumber \\ FMR_{Seq}(M,F) \cdot P(\bar{G},L) + IAPMR_{Seq}(M,F) \cdot P(\bar{G},\bar{L})
\end{eqnarray}
\normalsize

Since $P(\bar{G},*)/P(\bar{G})=P(*|\bar{G})$, we can rewrite Eq. \ref{eq:finalModel} as:
\small
\begin{eqnarray}
\label{eq:finalModelRED}
AR_{Seq}(M,F)=\nonumber\\ 
GAR_{Seq}(M,F) \cdot P(G) + GFMR_{Seq}(M,F; w) \cdot P(\bar{G})
\end{eqnarray}
\normalsize

Where $GFMR_{Seq}$ is for Global FMR (see also Ref. \cite{marcel2019HOA}, p. 471):

\small
\begin{eqnarray}
\label{eq:finalModel2}
GFMR_{Seq}(M,F) =\nonumber\\FMR_{Seq}(M,F) \cdot P(L|\bar{G}) + IAPMR_{Seq}(M,F) \cdot P(\bar{L}|\bar{G}) =\nonumber\\FMR_{Seq}(M,F) \cdot (1-w) + IAPMR_{Seq}(M,F) \cdot w
\end{eqnarray}
\normalsize
Worth noting, the term $w$ of Eq. \ref{eq:finalModel2} is also reported as a parameter, $\omega$, in Ref. \cite{marcel2019HOA}, p. 472, Eq. 20.5, with the following definition:
\textit{``$\omega$ denotes the relative cost of presentation attacks with respect to
zero-effort impostors''}. We have just proven that it is not a mere parameter adjusted to weight the attack's acceptance rate but corresponds to $P(\bar{L}|\bar{G})$, the prior probability of a presentation attack, according to the terminology adopted.
 
Eqs. \ref{eq:finalModelRED}-\ref{eq:finalModel2} define our simulator. The ROC curve of the sequential system is derived by considering the individual ROCs of the presentation attacks detector and the matcher. By acting on $w$, and the PAD's operational point $BPCER=p\%$ or $APCER=p\%$, the designer may depict several possible scenarios and decide whether the $GFMR_{Seq}$ is better than that of the individual matcher. By recalling that a PAD is tailored over the specific sensor, the simulator helps the designer select the most appropriate technology for the final application's security level.

\section{Experiments and Simulations}
\label{sec:exper}

\subsection{Datasets}
\label{subsec:protocol}

\begin{figure}
\centering     
\subfigure[]{\label{fig:a}\includegraphics[width=28mm]{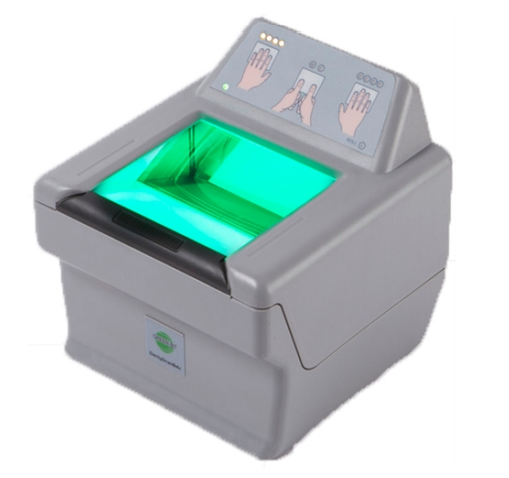}}
\subfigure[]{\label{fig:b}\includegraphics[width=28mm]{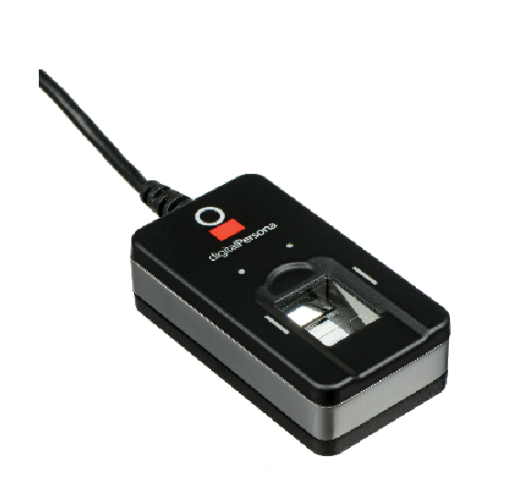}}
\subfigure[]{\label{fig:b}\includegraphics[width=22mm]{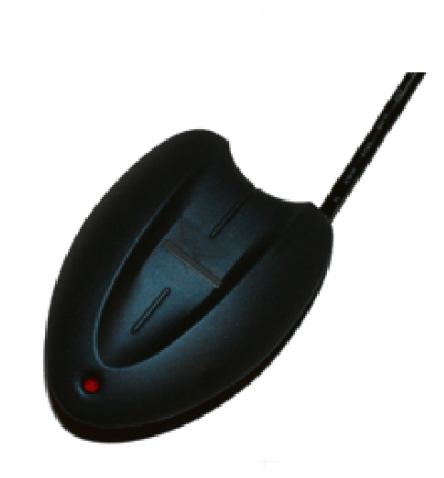}}
\caption{Sensors adopted in LivDet 2017 and 2019 editions: GreenBit (a), DigitalPersona (b) and Orcanthus (c).}
\label{fig:sensors}
\end{figure}

In the following experimental analysis, we utilized LivDet 2017 \cite{mura2015livdet} and LivDet 2019 \cite{orru2019livdet} datasets.
Both datasets consist of live and spoof fingerprint images from three different devices, two optical, GreenBit and Digital Persona, and a thermal swipe, Orcanthus (Figure \ref{fig:sensors}). The detailed characteristics of the sensors are shown in Table \ref{table:sensors}. 

\begin{table*}[ht]
\centering
\caption{Device characteristics for LivDet 2017 and LivDet 2019 datasets.}
\resizebox{\textwidth}{!}{%
\begin{tabular}[t]{ | l | l | c | c | c | c |}
\hline
\textbf{Scanner} & \textbf{Model} & \textbf{Resolution [dpi]} & \textbf{Image Size [px]} & \textbf{Format} &\textbf{Type} \\ \hline
Green Bit & DactyScan84C & 500 & 500x500 & BMP & Optical \\ \hline
Orcanthus & Certis2 Image & 500 & 300x$n$  & PNG & Thermal swipe \\ \hline
Digital Persona & U.are.U 5160 & 500 &252x324 & PNG & Optical \\ \hline
\end{tabular}}
\label{table:sensors}
\end{table*}

The spoof images of the LivDet 2017 and LivDet 2019 datasets were collected using the cooperative method. The materials used in the train set are different with respect to the test set as reported in Table \ref{table:datasetComposition2017} and Table \ref{table:datasetComposition2019}.

\begin{table*}[ht]
\centering
\caption{Composition of the LivDet 2017 dataset.}
\resizebox{\textwidth}{!}{%
\begin{tabular}{|c|c|c|c|c|c|c|c|c|c|c|c|c|}
\hline
                 & \multicolumn{4}{c|}{\textbf{Train}}      & \multicolumn{4}{c|}{\textbf{Test}}\\ \hline
\textbf{Dataset} & Live & Wood Glue & Ecoflex & Body Double & Live & Gelatine & Latex & Liquid Ecoflex\\ \hline
Green Bit        &1000&400&400&400&1700&680&680&680\\ \hline
Orcanthus        &1000&400&400&400&1700&680&658&680\\ \hline
Digital Persona  &999&400&400&399&1700&679&670&679\\ \hline
\end{tabular}%
}
\label{table:datasetComposition2017}
\end{table*}

\begin{table*}[ht]
\centering
\caption{Composition of the LivDet 2019 dataset.}
\resizebox{\textwidth}{!}{%
\begin{tabular}{|c|c|c|c|c|c|c|c|c|c|c|c|c|c|c|}
\hline
                 & \multicolumn{6}{c|}{\textbf{Train}}      & \multicolumn{4}{c|}{\textbf{Test}}\\ \hline
\textbf{Dataset} & Live & Wood Glue & Ecoflex & Body Double & Latex & Gelatine & Live & Mix 1 & Mix 2 & Liquid Ecoflex\\ \hline
Green Bit        &1000&400&400&400& - &-&1020&408&408&408\\ \hline
Orcanthus        &1000&400&400&400&-&-&990&384&308&396\\ \hline
Digital Persona  &1000&250&250&-&250&250&1019&408&408&408\\ \hline
\end{tabular}%
}
\label{table:datasetComposition2019}
\end{table*}

\subsection{Protocol}
The first necessary step to corroborate the thesis expressed by Eqs. \ref{eq:AR_LM} consists of measuring the differences, in terms of performance, between a real sequential integrated system and our simulated scenario.
If they are comparable, we can profitably employ the model to investigate the actual aftermath that a liveness detector module can bring to a fingerprint verification system before implementing the fusion itself.

To this purpose, we considered several PAD systems based both on hand-crafted features and deep learning methods. Specifically, we exploited all the algorithms submitted to LivDet 2017 and 2019 competitions by testing them on relative edition datasets: while in the 2017 edition the solutions adopted are equally distributed between deep learning and hand-crafted algorithms, in 2019, almost all detectors are based on deep learning approaches.

Subsequently, we designed the following experimental protocol\footnote{In order to guarantee the repeatability of experiments, the authors are willing to share the source code they used.}:

\begin{enumerate}

\item We computed the liveness score and match score (using the standard Bozorth3 matcher) on the relative test sets.
\item We computed the individual ROC curves for both PAD and verification system.
\item We estimated the theoretical ROC curve of the sequential integrated system by applying Eqs.\ref{eq:AR_LM}. According to our findings, there is no substantial difference between the $Match \rightarrow Liveness$ and the $Liveness \rightarrow Match$ system. Since the final decision can also be configured as an AND-like boolean one, that is, the pattern is finally accepted when $F$ and $M$ events are both $True$. In our architecture, the liveness module precedes the matcher.
\item We computed the performance of the system separately, without the help of Eqs. \ref{eq:AR_LM}. In other words, we computed the experimental ROC curves according to the standard design approach.
\end{enumerate}

Among all possible operational points of the PAD module investigated, we focused in particular on two of them:
\begin{itemize}
  \item $APCER(s^*_F)=0.01$, where only $1\%$ of presentation attacks can be misclassified. We indicate it as $APCER_{01}$ operational point.
  \item $BPCER(s^*_F)=0.01$, where $1\%$ of live samples is incorrectly classified. We refer to it as $BPCER_{01}$ operational point.
\end{itemize}

We choose the selected ones since they represent two case-studies quite extreme: the first might exemplify a context where it is necessary to tolerate very few attacks, due to security constraints; the second case is typical of services addressed to a large number of different users, where instead it is supposed that the probability of presentation attacks is lower over time, and it is much more important that no users are "blocked" by the PAD module.

 \subsection{Validation}
\label{subsec:results}

The following experimental analysis aims to point out the model's reliability in predicting a real sequential system.
For all the chosen detectors, we computed the absolute difference between significant indexes (FMR, GAR and IAPMR) estimated by Eqs. \ref{eq:AR_LM} and those obtained through the standard design approach, at the selected operational points.
For the sake of space, we then extrapolated from the results a set of statistical parameters to show the estimation error, expressed in percentage points, introduced by the model, and we summarized them with the help of box plots. Since the purpose of this Section is not to underline the performance differences over the sensors or to assess the best PAD, for both editions of LivDet, we considered a global estimation error, namely one error per acceptance rate. We report the mean and standard deviation of such error in Table \ref{table:meanBP}.

\begin{table}[!t]
\begin{center}
\caption{LivDet 2017 and 2019 datasets: Mean and standard deviation of the absolute difference of FMR, GAR, and IAPMR between a standard and an estimated scenario for the two investigated operational points. Reported values are not fractional.}
\label{table:meanBP}

\begin{tabular}[t]{ | c | c | c || c |}
\hline
\multicolumn{2}{|c|}{} & LivDet 2017 & LivDet 2019 \\
\hline\hline
\multirow{3}{*}{{$APCER_{01}$}} 
& FMR & 0.0265 $\pm$ 0.023      & 0.0206 $\pm$ 0.023   \\ \cline{2-4}
& GAR & 0.9376 $\pm$ 0.388      & 0.2911 $\pm$ 0.265  \\ \cline{2-4}
& IAPMR & 0.0254 $\pm$ 0.021     & 0.0475 $\pm$ 0.136 \\ 
\hline\hline
\multirow{3}{*}{$BPCER_{01}$ }
& FMR   & 0.0094 $\pm$ 0.011    & 0.0060 $\pm$ 0.008  \\ \cline{2-4}
& GAR   & 0.1532 $\pm$ 0.046    & 0.1330 $\pm$ 0.092  \\  \cline{2-4}
& IAPMR  & 0.1910 $\pm$ 0.150    & 0.0855 $\pm$ 0.133 \\ \hline

\hline

\end{tabular}
\end{center}
\label{tab:sensors}
\end{table}

We present in the following four figures that are aimed to roughly show that the AR curves predicted by the model are strongly similar to the ones obtained by a real set of probes submission to the system. 

\begin{figure}
\centering     
\subfigure[]{\label{fig:a}\includegraphics[width=.45\textwidth]{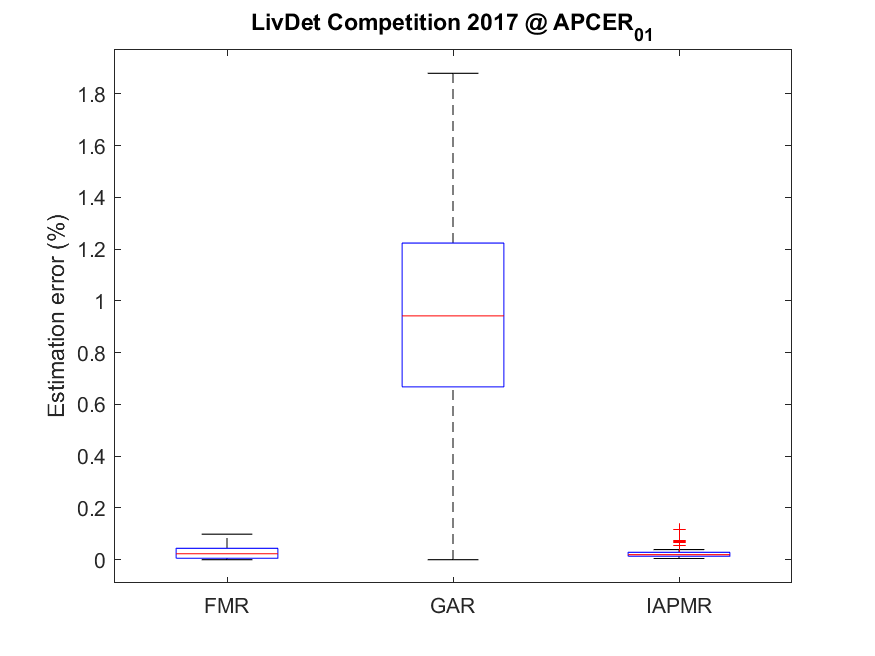}}
\subfigure[]{\label{fig:b}\includegraphics[width=.45\textwidth]{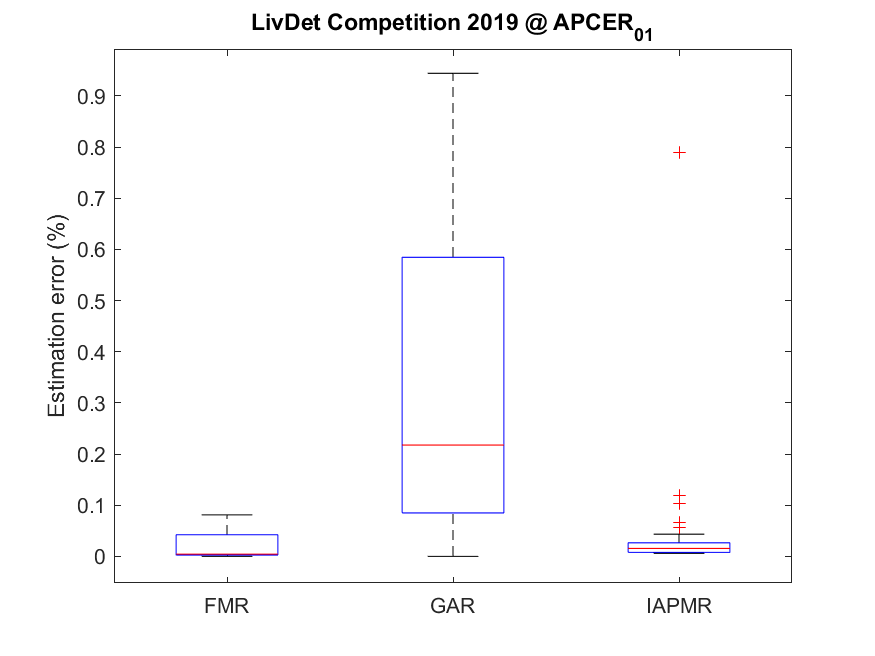}}
\caption{Box plots of the absolute difference of FMR, GAR, and IAPMR between a standard and an estimated scenario for the $APCER_{01}$ operational point in LivDet 2017 (a) and 2019 (b).}
\label{fig:livFPR}
\end{figure}

The former set of images (Fig. \ref{fig:livFPR}) shows the systems’ performances at the $APCER_{01}$ operational point, namely when using a very precautionary liveness threshold. Firstly, we notice that the two LivDet datasets are similar in terms of error distribution: the interquartile range (IQR) of genuine acceptance rate (GAR) distribution, that is, the difference between the third quartile (Q3) and the first one (Q1) and graphically, the total box length, is relatively more extensive than that of IAPMR and FMR distributions. In other words, the values in GAR distribution are characterized by a more significant variability, particularly for the 2019 edition, while IAPMR and FMR observations appear to be more consistent. By observing the median’s bottom position in 2019 distributions, it is apparent that they present a positive skew; namely, most of the observations are concentrated on the low end of the scale. 
Furthermore, another critical point to highlight is the presence of outliers. By definition, if a value is outside of the $Q3 + 1.5\cdot IQR$ or the $Q1 - 1.5\cdot IQR$ range, that value will be considered an outlier. There are many strategies for dealing with outliers in data, depending on the application and dataset. 
Fig. \ref{fig:livFPR} shows outliers only on IAPMR error distribution and analyzing their value, we can state that they have no statistical significance for the simulation purposes of our model.

\begin{figure}
\centering     
\subfigure[]{\label{fig:a}\includegraphics[width=.45\textwidth]{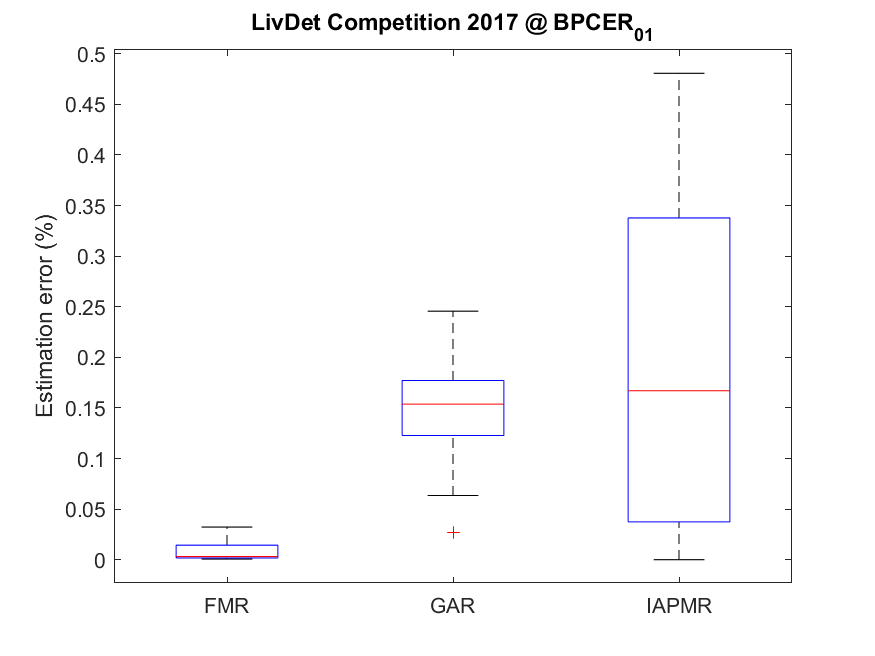}}
\subfigure[]{\label{fig:b}\includegraphics[width=.45\textwidth]{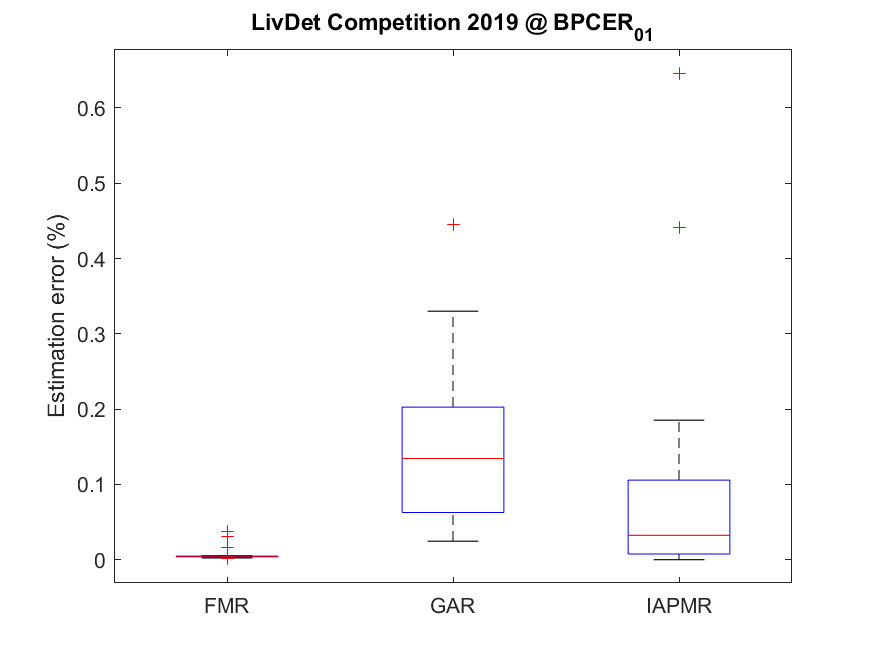}}
\caption{Box plots of the absolute difference of FMR, GAR, and IAPMR between a standard and an estimated scenario for the $BPCER_{01}$ operational point in LivDet 2017 (a) and 2019 (b).}
\label{fig:livTPR}
\end{figure}

Fig. \ref{fig:livTPR} shows an example of what happens when a very relaxed liveness threshold is set, namely, when the liveness detector is working at $BPCER=1\%$. In this case, we assist at a considerable growth of IAPMR estimation error, in favour of a GAR and FMR distribution decrease. This was partly expected, since the integrated system performances with a tolerant PAD tend to those of the matching system alone. Overall, the data asymmetry is less marked. Thus, the probability of getting estimation errors is higher than in the previous instance. Nevertheless, the maximum range is smaller since the largest non-outlier for both editions is relatively lower. Another difference is represented by outliers’ appearance also in FMR and GAR error distributions, although they do not mean a severe threat to the goodness of fit, as formerly stated. 

In summary, this investigation's findings suggest that the predicted performances are consistent with the experimental curves. The maximum error value of this data collection was made by a 2017 liveness detector in GAR distribution at the $APCER_{01}$ operational point, and it is approximately equal to 1.88. In other words, the worst dissimilarity between our model and the real curve is, on average, lower than $2\%$.
Therefore, we can conclude that our model effectively simulates the sequential combination of a PAD and a fingerprint verification system with an acceptable degree of tolerance.

\subsection{Simulations: are we ready for integration?}

Once the experimental analysis confirmed the estimate's validity, the proposed simulator allows us to focus on verifying the improvement achievable by embedding state-of-the-art PAD modules into a state-of-the-art fingerprint verification one. 

In the previous Section, we designed a model for the GFMR index (Eq. \ref{eq:finalModel2}), defined as a weighted sum of FMR and IAPMR through the term $w$. Thanks to this term, which represents the PA probability, it is possible to evaluate the system's response, with and without liveness detection embedded, by Eqs. \ref{eq:AR_LM}.
By way of example, we show the ROC curves plotting the benefits of having an intrinsically “secure” fingerprint verification system when $w=[0.00,0.75]$. Choosing such a large, though unlikely, range allows us not only to test different security scenarios but also to show PADs' behavior in the worst possible case.
To this aim, we offer two sets of plots: in the first one, we compare the integrated with the corresponding individual system equipped with the standard matcher Bozorth3; in the other, the comparison is carried out by adopting the top-level matcher Verifinger 12.

We analyze all the algorithms submitted to LivDet 2017 and 2019 competitions. Nevertheless, for the sake of space, we propose only the results of the winning algorithm submitted to LivDet 2019 edition, namely, the algorithm named "PADUnkFv" \cite{padunk}, since its behavior is representative of the majority of the presented algorithms.
We tested all the datasets captured in that edition and the previous one, in order to set cross-dataset and cross-material experiments.
From this investigation, we identified two critical cases for which the integration causes significant degradation of overall personal verification performance: one is due to the acquisition sensor and the other is related to the PAI material.\newline

\subsubsection{Sensor dependent analysis}
\label{subsec:sensordep}

\begin{figure*}[!h]
\begin{center}
    \centering 
    \subfigure[]{
    \includegraphics[width=.45\textwidth]{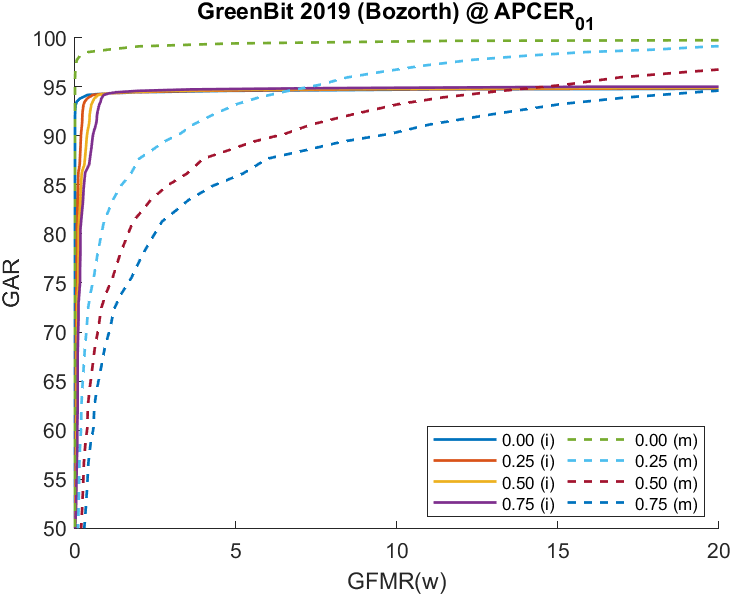}}
    \subfigure[]{
    \includegraphics[width=0.45\textwidth]{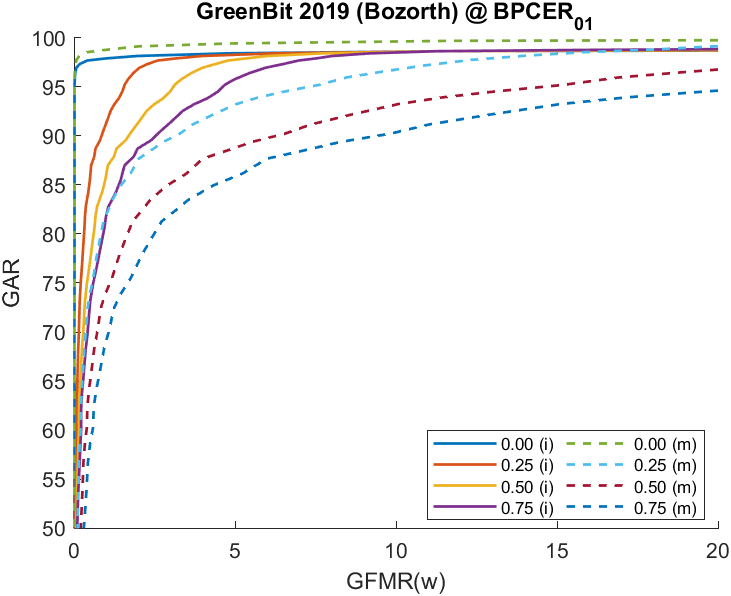}}
    \\
    \subfigure[]{
    \includegraphics[width=0.45\textwidth]{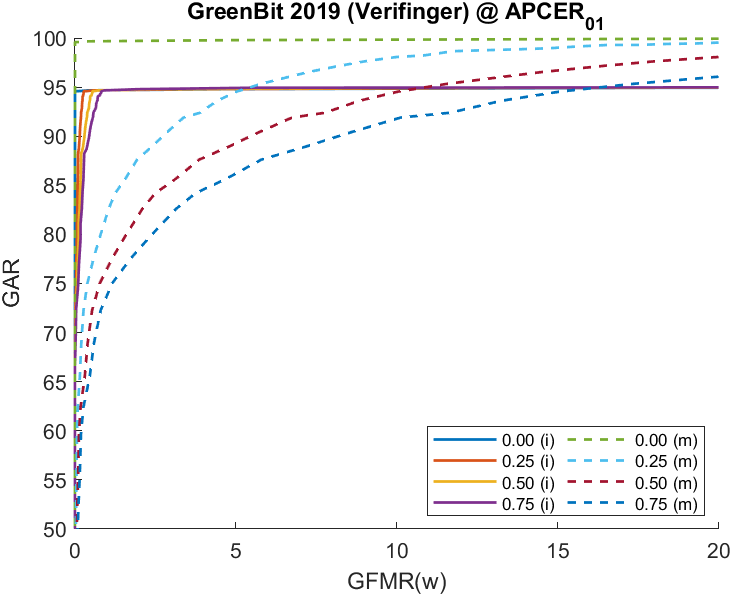}}
    \subfigure[]{
    \includegraphics[width=0.45\textwidth]{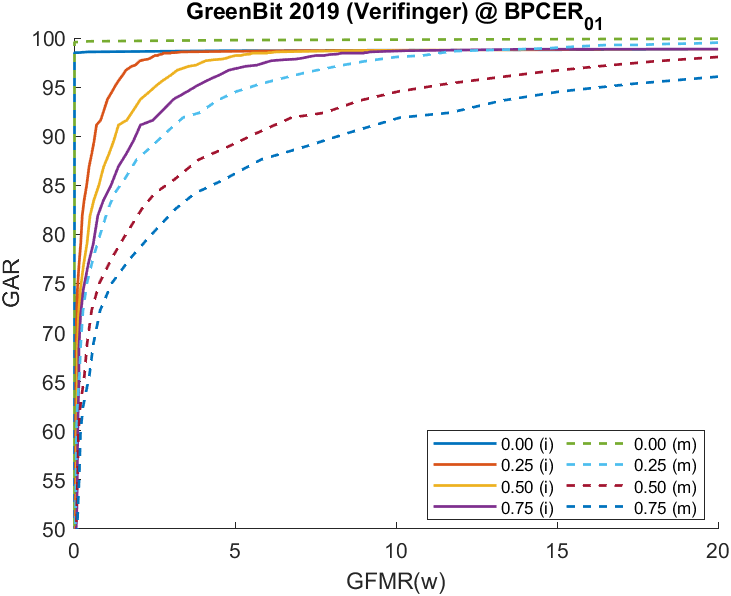}}

\caption{GreenBit dataset from LivDet 2019.  Comparison between GROCs for an integrated (solid line) and individual (dashed line) matching system equipped with Bozorth3 (a,b) and Verifinger 12 (c,d), varying the presentation attacks probability $w$. Both operational points are reported for each matcher.}
\label{fig:GbExFPR}
\end{center}
\end{figure*}

\begin{figure*}[!h]
   \centering
   \subfigure[]{\includegraphics[width=.45\textwidth]{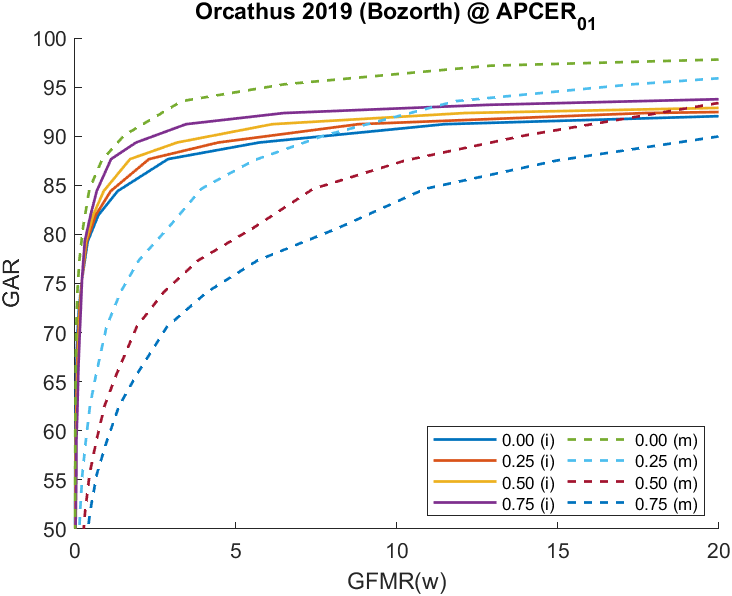}}
      \centering
   \subfigure[]{\includegraphics[width=.45\textwidth]{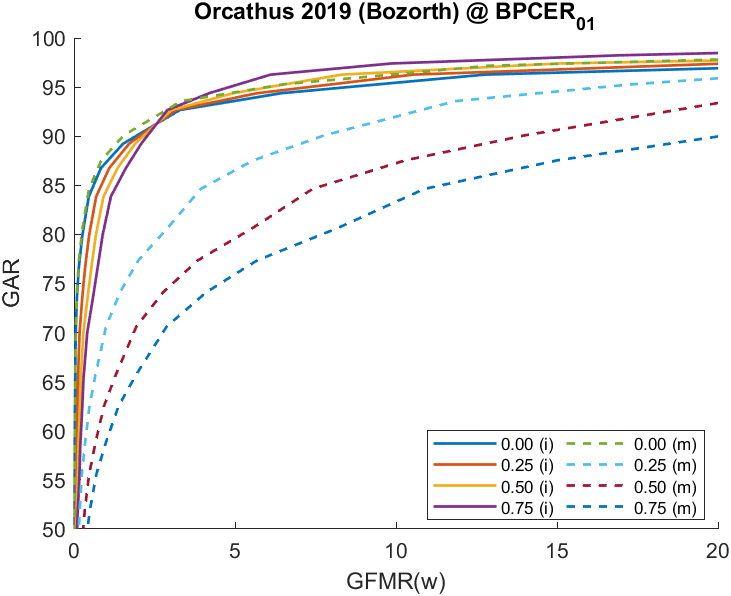}}\\
      \subfigure[]{\includegraphics[width=.45\textwidth]{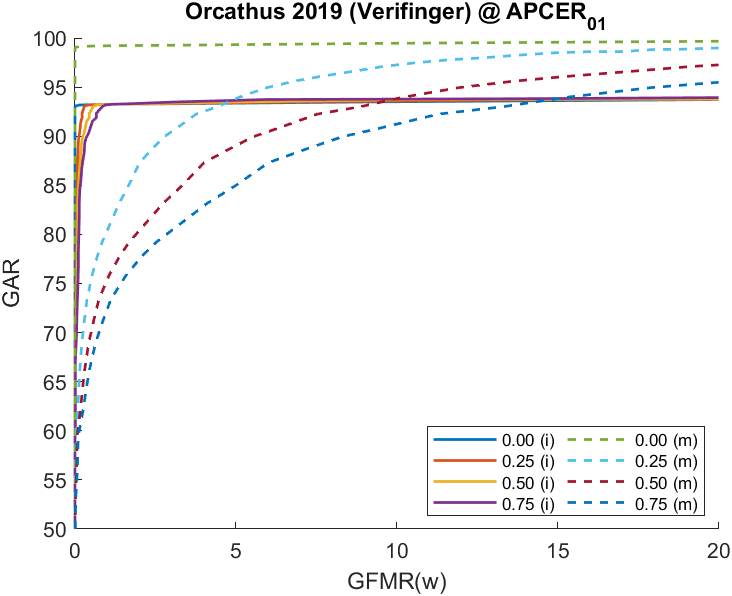}} 
   \subfigure[]{\includegraphics[width=.45\textwidth]{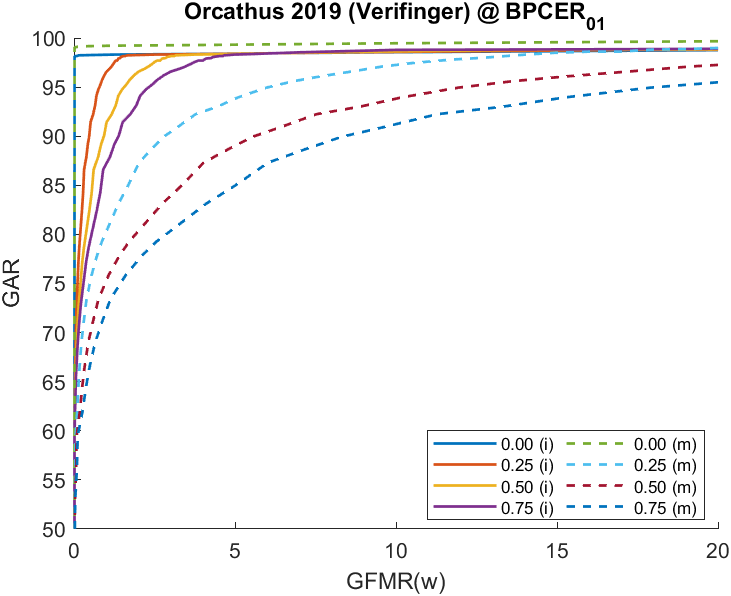}}
\caption{Orcanthus dataset from LivDet 2019. Comparison between GROCs for an integrated (solid line) and individual (dashed line) matching system equipped with Bozorth3 (a,b) and Verifinger 12 (c,d), varying the presentation attacks probability $w$. Both operational points are reported for each matcher.}
\label{fig:OrExFPR}
\vspace{-10pt}
\end{figure*}

The following results offer a clear overview of the integration in the two extreme cases examined in this work. 
When the system is adjusted to the $APCER=1\%$ operational point, which can be considered the most crucial working point since we are assuring that a tiny percentage of presentation attacks can be tolerated,  it is evident that the integration usefulness depends entirely on the probability of attack $w$.
If the PA is very high ($w>0.50$), it is advisable to use the integrated system. Vice versa, putting a PAD becomes no longer convenient since there is a loss in terms of GAR;  this loss is intrinsic to this kind of fusion and depends on the goodness of the liveness module. In this case, the designer could avoid the use of the PAD or could use its response as a sort of ”warning” information during the system’s operations, or the threshold could be relaxed to obtain a better performance on genuine users, whereas, however, misclassifying more fake fingerprints.
On the other hand, when the genuine acceptance is the first care ($BPCER = 1 \%$), the performances are much more balanced, and the integrated system does not exhibit a notable $APCER$ value increase by strongly reducing the IAPMR as $w$ rises.
The common aspects highlighted by Figures \ref{fig:GbExFPR}-\ref{fig:OrExFPR} are mainly two:
 
\begin{enumerate}
\item A system with liveness detector embedded is more robust to $w$ variations than a simple matcher. Graphically, we can notice that the curves dispersion is nearly null, in other words, the performance does not decay if $w$ increases, as is the individual case. This means that the PAD is working correctly and blocking attack attempts from fingerprint forgers. 
\item Although Verifinger 12 is a top-level fingerprint verification algorithm, compared with Bozorth3, 
the absence of substantial performance differences when considering their integration with the PAD suggests that \textit{this} is the leading ``actor'' to the global system's effectiveness. We can therefore neglect the dissimilarities among the two matchers and focus the attention on the PAD's characteristics.
\end{enumerate}

\begin{figure*}[!h]
   \centering
      \subfigure[]{\includegraphics[width=.45\textwidth]{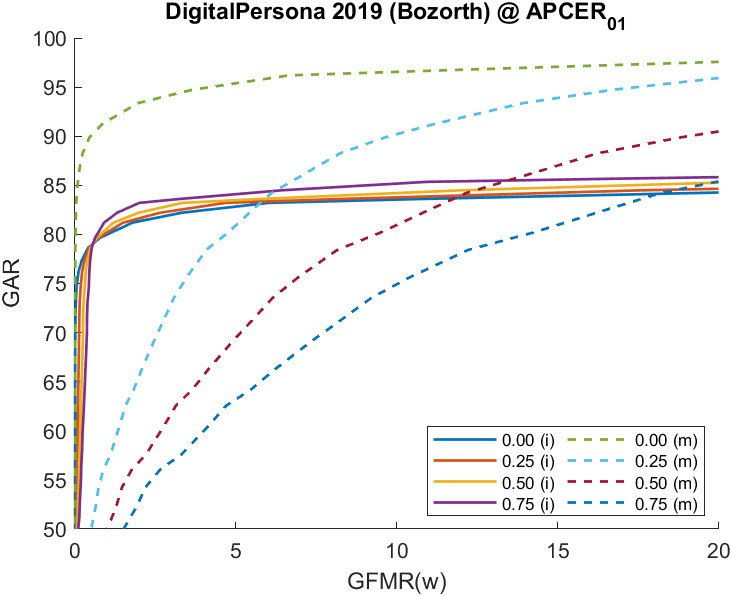}}
   \subfigure[]{\includegraphics[width=.45\textwidth]{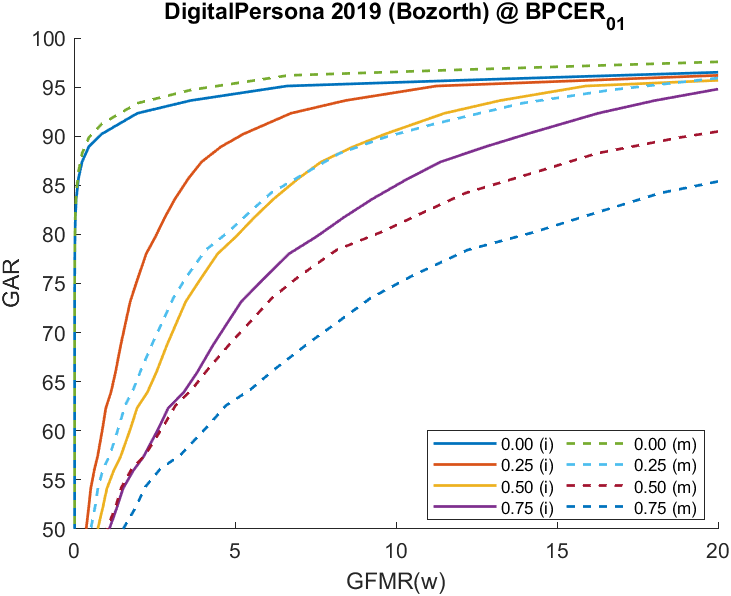}}\\
   \subfigure[]{\includegraphics[width=.45\textwidth]{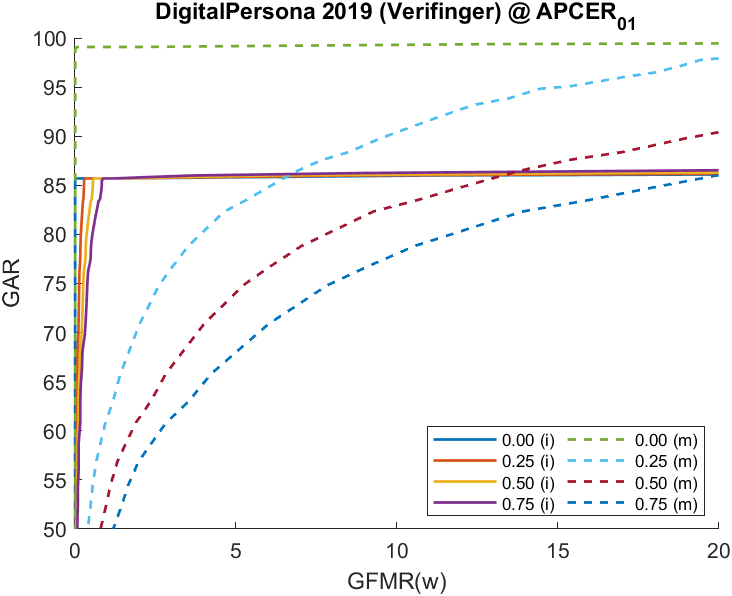}}
   \subfigure[]{\includegraphics[width=.45\textwidth]{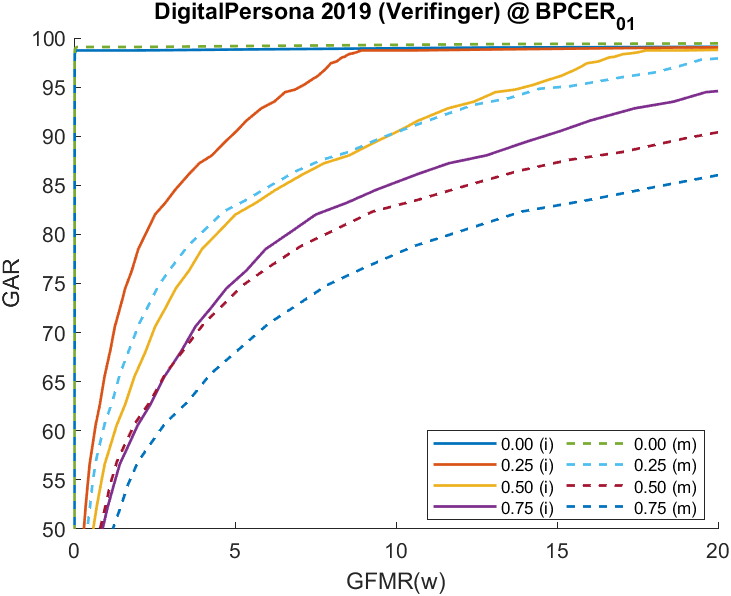}} 
\caption{DigitalPersona dataset from LivDet 2019. Comparison between GROCs for an integrated (solid line) and individual (dashed line) matching system equipped with Bozorth3 (a,b) and Verifinger 12 (c,d), varying the presentation attacks probability $w$. Both operational points are reported for each matcher.}
\label{fig:DP_verif_FPR_TPR}
\vspace{-10pt}
\end{figure*}

Let us now consider the fingerprint verification system working on the Digital Persona sensor. Figure \ref{fig:DP_verif_FPR_TPR}a-c shows the PAD performance, achieved at $APCER_{01}$ operational point. We immediately notice that the performance drop is considerably higher if compared to the other competition sensors.
This is significant evidence: wherever the meta-designer chooses to operate in a very conservative operational point, he/she can decide to act in advance on the system parameters to improve the performances.

It can be assumed that this behavior is due to the fact that the acquisition surface is significantly reduced when compared to that of the GreenBit sensor, as reported in Table \ref{table:sensors} (the Orcanthus sensor has a different acquisition technology), and this represents an obstacle in capturing the defects that may appear in the fake fingerprint edge and that could facilitate the PA detection. To support this thesis, we report the following Figure \ref{fig:meanFPR_sensors} representing the average attack presentation classification error rate computed on all LivDet 2019 algorithms: the shape of the $APCER$ curve is more relaxed in the case of the DigitalPersona sensor, which means that the percentage of false positives (i.e., the fake fingerprints classified as alive) is greater under the same threshold. For this reason, the $APCER_{01}$ operational point corresponds to a value of the liveness threshold such as to lead to misclassification of a significant number of live fingerprints.

\begin{figure}[tb]
   \centering
   \includegraphics[width=.45\textwidth]{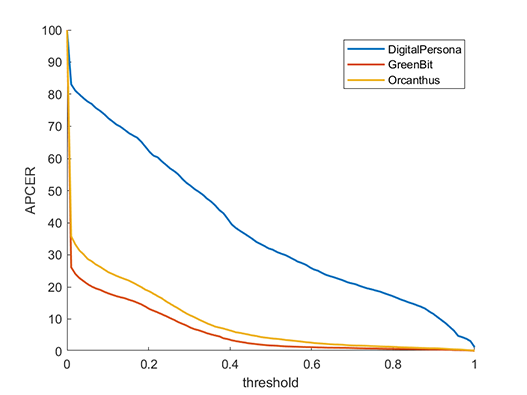}
   \caption{Average Attack Presentation Classification Error Rate (APCER) compared in the three sensors of the LivDet 2019 edition. A slow decay trend characterizes the case of the DigitalPersona dataset if compared to the other sensors.}
\label{fig:meanFPR_sensors}
\vspace{-10pt}
\end{figure}

On the other hand, when the PAD is working at $BPCER_{01}$ operational point (Fig. \ref{fig:DP_verif_FPR_TPR}\{b,d\}), the integrated system  performance improves and gains a fair degree of robustness as the probability of attack increases. As we pointed out in the previous section, the operational points at which both system operate impact deeply on the final performance, since we have basically the product of the individual error/detection rate.

\subsubsection{Material dependent analysis}
\label{subsec:materialdep}

Another simulation ability consists of showing us the impact of sensitive materials on the system, where they exist.
It is commonly acknowledged that the liveness detector reacts differently depending on the spoof material \cite{marasco2014survey, marcel2014handbook}. How does this impact on the integrated system? How do GAR and GFMR change?
If the PAD behavior is not consistent with all materials, then an attack with a certain material may completely change the design expectations.
We remember that our purpose is not to understand what system attains the best performance but to point out the information that a designer can exploit in the early project phase. This allows us to draft the design guidelines by mean of the proposed simulator.

\begin{figure*}[!h]
   \centering
   \subfigure[]{\includegraphics[width=.45\textwidth]{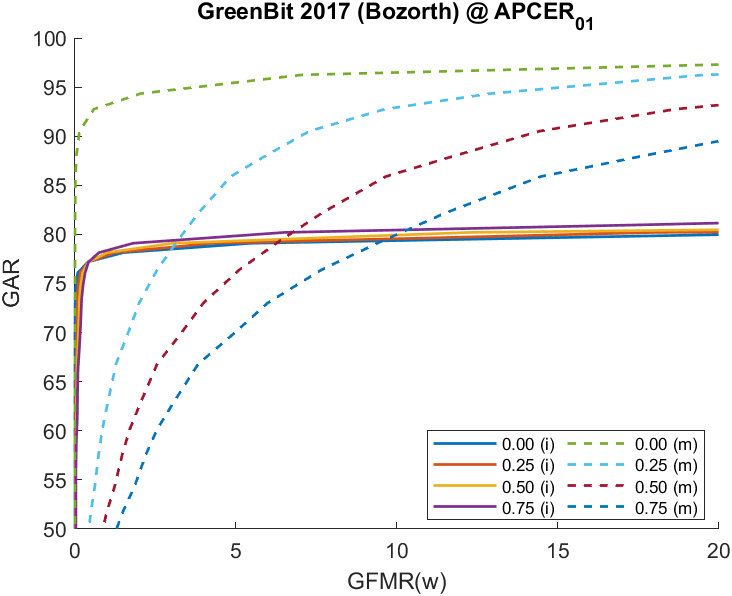}}
     \subfigure[]{\includegraphics[width=.45\textwidth]{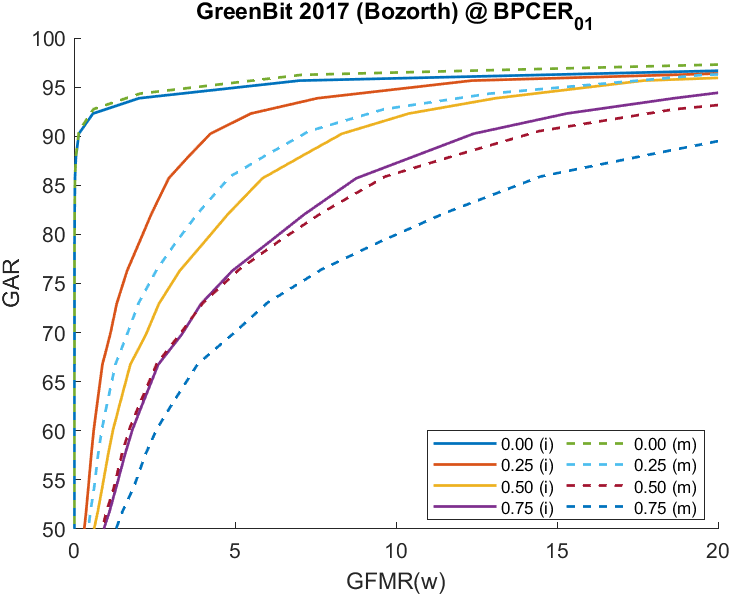}}\\
    \subfigure[]{\includegraphics[width=.45\textwidth]{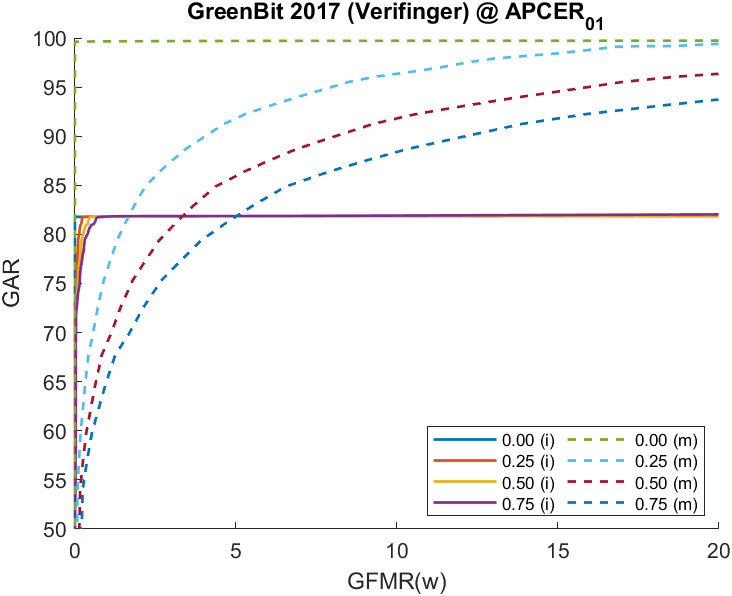}}
   \subfigure[]{\includegraphics[width=.45\textwidth]{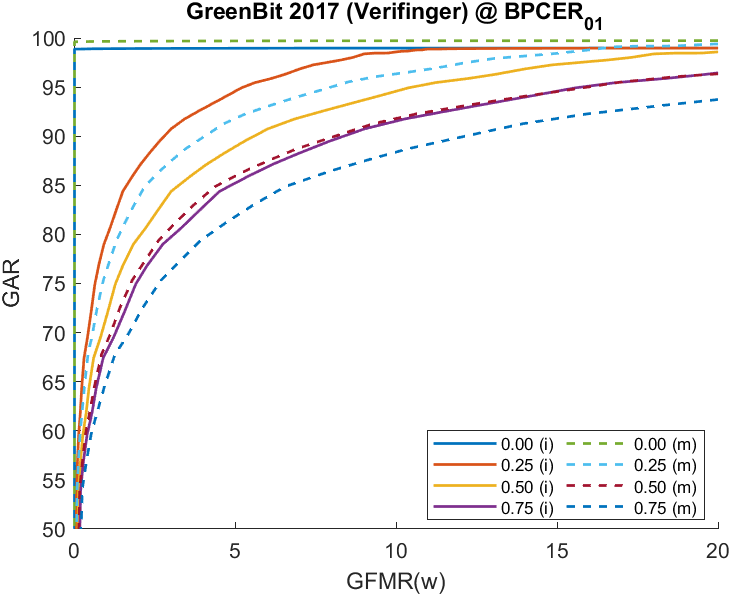}}
\caption{GreenBit dataset from LivDet 2017. Comparison between GROCs for an integrated (solid line) and individual (dashed line) matching system equipped with Bozorth3 (a,b) and Verifinger 12 (c,d), varying the presentation attacks probability $w$. Both operational points are reported for each matcher.}
\label{fig:Gb17ExFPR}
\vspace{-10pt}
\end{figure*}

\begin{figure*}[!h]
   \centering
   \subfigure[]{\includegraphics[width=.45\textwidth]{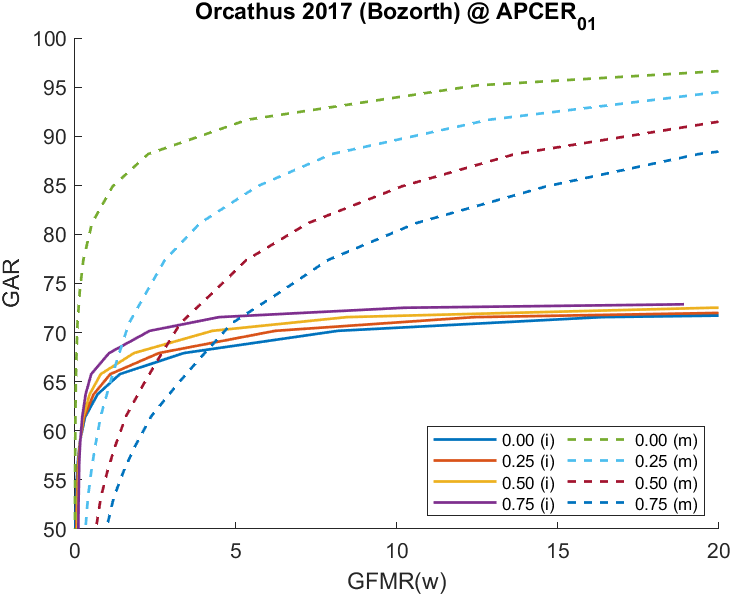}}
   \centering
   \subfigure[]{\includegraphics[width=.45\textwidth]{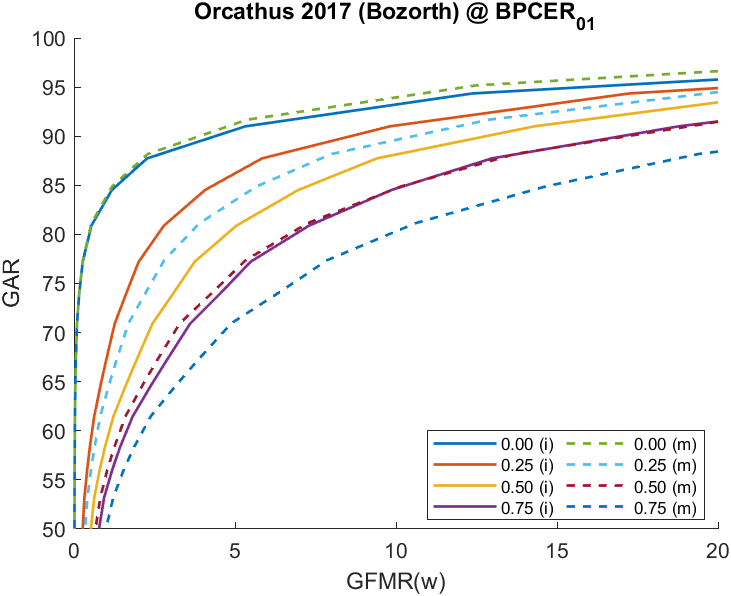}}\\
      \subfigure[]{\includegraphics[width=.45\textwidth]{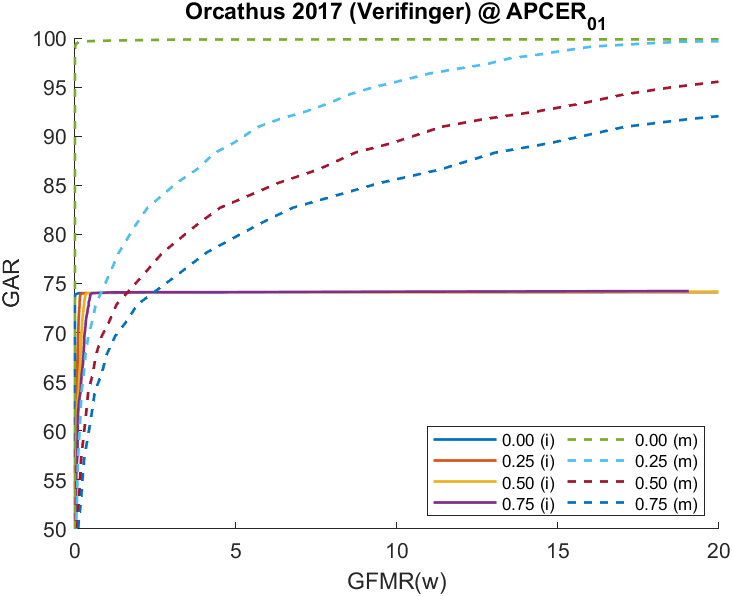}} 
   \subfigure[]{\includegraphics[width=.45\textwidth]{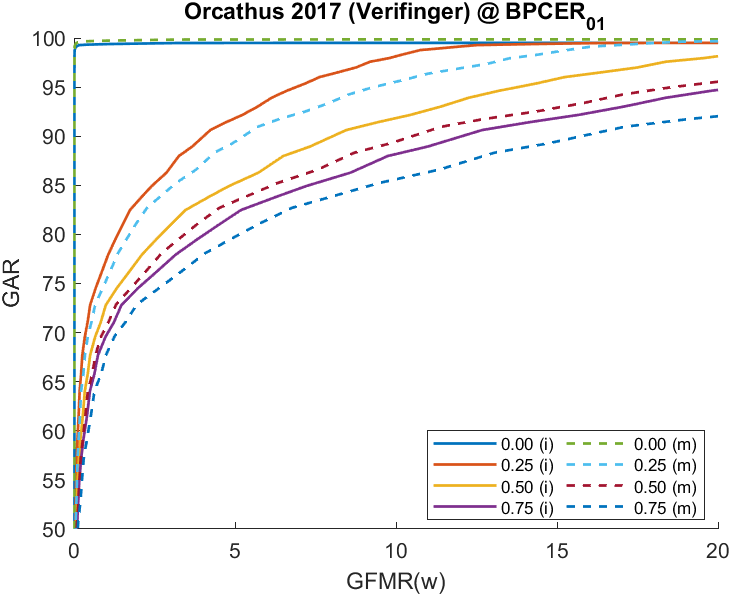}}
\caption{Orcanthus dataset from LivDet 2017. Comparison between GROCs for an integrated (solid line) and individual (dashed line) matching system equipped with Bozorth3 (a,b) and Verifinger 12 (c,d), varying the presentation attacks probability $w$. Both operational points are reported for each matcher.}
\label{fig:Or17ExFPR}
\end{figure*}

The three test sets of LivDet2017 \cite{mura2015livdet}, as reported in Table \ref{table:datasetComposition2017} and \ref{table:datasetComposition2019}, differ in both the fingerprints number and composition, but the training set is the same regarding the GreenBit and Orcanthus sensors.
This allowed us to draw conclusions on the critical or favorable material classes for a specific detector type.
For this purpose, similar to the case described in the previous section, we tested the LivDet 2019 winner on the LivDet 2017 datasets. Thus, we noticed a significant drop in the classification accuracy, compared to the respective 2019 datasets, when the liveness threshold is such that $FPR = 1 \%$ (Fig. \ref{fig:Gb17ExFPR}\{a,c\} and \ref{fig:Or17ExFPR}\{a,c\}).

Since we are analyzing the same acquisition sensors, in this case the cause of this behavior can only be due to the material diversity used in the two competitions.
Therefore, a further analysis was carried out at the level of classification on the individual materials.

\begin{figure}[!h]
   \centering
   \includegraphics[width=0.45\textwidth]{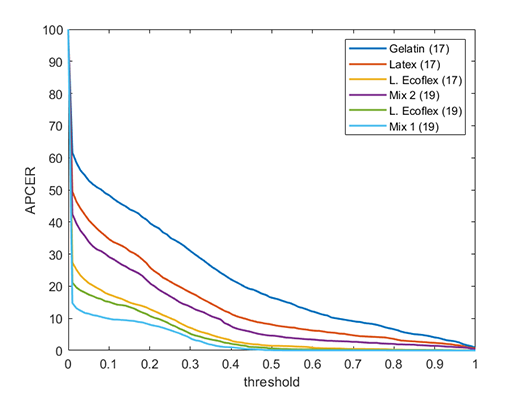}
   \caption{Average APCER compared for the GreenBit sensor of LivDet 2017 and 2019. The "Gelatin" material used in LivDet 2017 originates fake fingerprints harder to classify. }
\label{fig:meanFPR_materials}
\vspace{-10pt}
\end{figure}

Figure \ref{fig:meanFPR_materials} clearly shows the difference in the two cases above: the LivDet2017's gelatin-made spoofs are erroneously classified as alive fingerprints in a percentage higher than that of other materials. This means that the gelatin can reproduce the real fingerprint's characteristics with such an accuracy that an artifact of the kind can deceive the detector with a probability higher than others. 

Moreover, the percentage difference between the two indices related to common materials ("L. Ecoflex" and "Latex"\footnote{The "Mix 2" material in the 2019 competition is a mixture of Latex (70\%) and glue (30\%), therefore for comparative purposes, it can be approximated to the pure “Latex” used in the previous edition.}) is strongly reduced if compared to that between "Mix" and "Gelatine".

If a more relaxed threshold is selected as in $BPCER=1\%$ case, the integrated system performance tends to improve, especially when $w$ rises.

Therefore, the presentation attacks detector must be studied concerning critical materials before being selected. In general, we can state that there some materials still represent an open issue, as the best LivDet2019 detector cannot achieve a good performance on them. This is true for gelatin, which is, unfortunately, one of the most commonly available and cheapest materials.\newline

\subsubsection{PAD dependent analysis}
\label{subsec:PADdep}

A last investigation is proposed to corroborate the hypothesis about the PAD weight in the integrated system's efficiency. The graphs reported so far show that the performance between two sequential systems equipped with two different matching solutions are approximately equals, implicitly indicating that the crucial phase during the design phase lies in the liveness detector's choice.

\begin{table*}
\centering
\caption{LivDet 2019 dataset. Comparison of BPCER@1\%BPCER and APCER@1\%BPCER for the two most accurate liveness detectors of the LivDet 2019 competition.}
\resizebox{1\textwidth}{!}{%
\begin{tabular}{|c|c|c|c|c|c|c|c|} 
\cline{3-8}
\multicolumn{1}{c}{} &               & \multicolumn{2}{c|}{\textbf{GreenBit}}                                                                                                                      & \multicolumn{2}{c|}{\textbf{DigitalPersona}}                                                                                                                                                  & \multicolumn{2}{c|}{\textbf{Orcanthus}}                                                                                                                                                        \\ 
\hline
\textbf{Name}        & \textbf{Type} & \begin{tabular}[c]{@{}c@{}}\textbf{BPCER }\\\textbf{@1\% APCER}\end{tabular} & \begin{tabular}[c]{@{}c@{}}\textbf{APCER }\\\textbf{@1\% BPCER}\end{tabular} & \begin{tabular}[c]{@{}c@{}}\textbf{\textbf{BPCER}}\\\textbf{\textbf{@1\% APCER}}\end{tabular} & \begin{tabular}[c]{@{}c@{}}\textbf{\textbf{APCER}}\\\textbf{\textbf{@1\% BPCER}}\end{tabular} & \begin{tabular}[c]{@{}c@{}}\textbf{\textbf{BPCER}}\\\textbf{\textbf{@1\% APCER}}\end{tabular} & \begin{tabular}[c]{@{}c@{}}\textbf{\textbf{APCER}}\\\textbf{\textbf{@1\% BPCER}}\end{tabular}  \\ 
\hline
\textbf{PADUnkFv \cite{padunk}}    & Deep learning & 5.00\%                                                                       & 14.22\%                                                                      & 14.03\%                                                                                       & 40.95\%                                                                                       & 5.96\%                                                                                        & 5.88\%                                                                                         \\ 
\hline
\textbf{JLW\_LivDet} & Hand-crafted  & 0.39\%                                                                       & 0.33\%                                                                       & 26.67\%                                                                                       & 55.60\%                                                                                       & 3.23\%                                                                                        & 5.51\%                                                                                         \\
\hline
\end{tabular}}
\label{table:BPCERAPCER19}
\end{table*}

To explicitly confirm this conjecture, we present in the following some experimental evidence that testifies the behavior of two distinct PADs. In particular, we tested the LivDet 2019 winner mentioned above with the runner-up, the algorithm called "JLW\_LivDet". We summarize in Table \ref{table:BPCERAPCER19} the main differences between these two algorithms by reporting their performances in terms of BPCER@1\%APCER and APCER@1\%BPCER on all datasets of LivDet 2019 edition.

\begin{figure*}[!h]
   \centering
      \subfigure[]{\includegraphics[width=.45\textwidth]{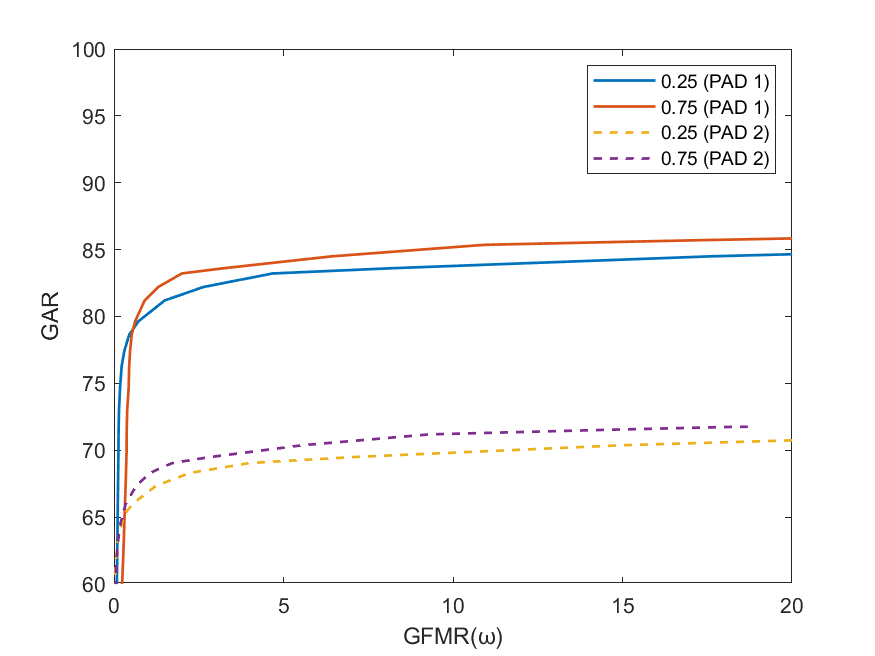}}
     \subfigure[]{\includegraphics[width=.45\textwidth]{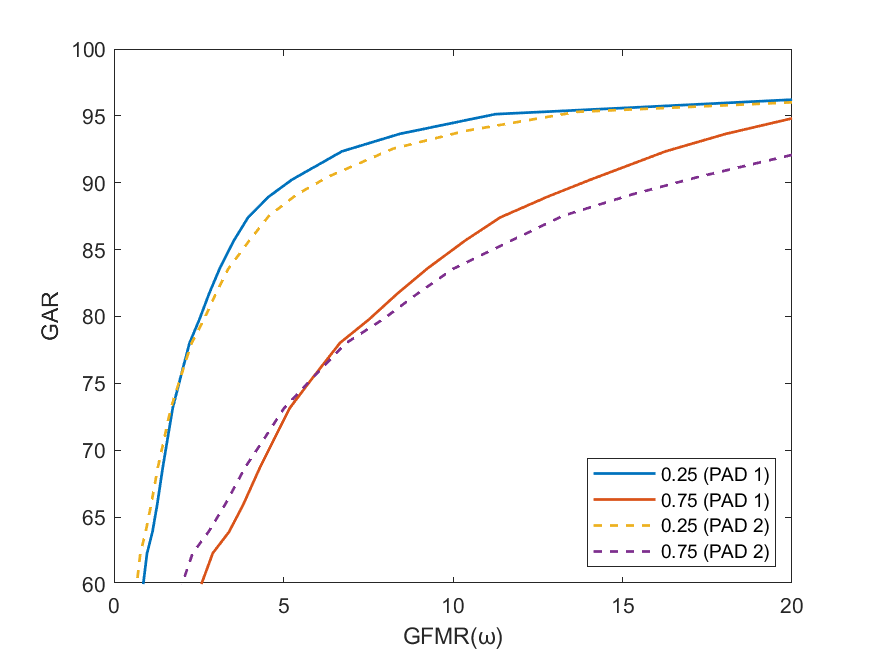}}\\
   \subfigure[]{\includegraphics[width=.45\textwidth]{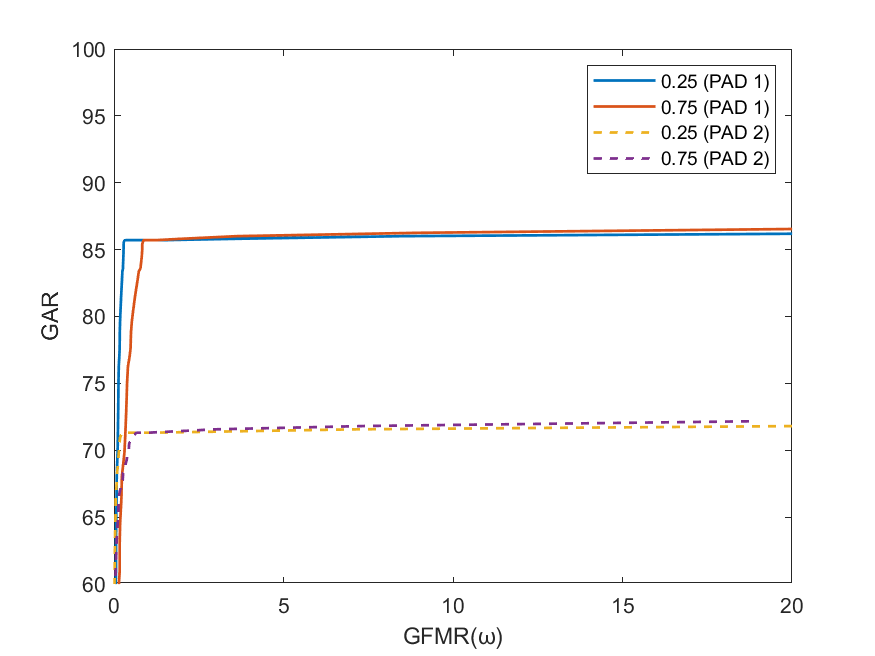}}
     \subfigure[]{\includegraphics[width=.45\textwidth]{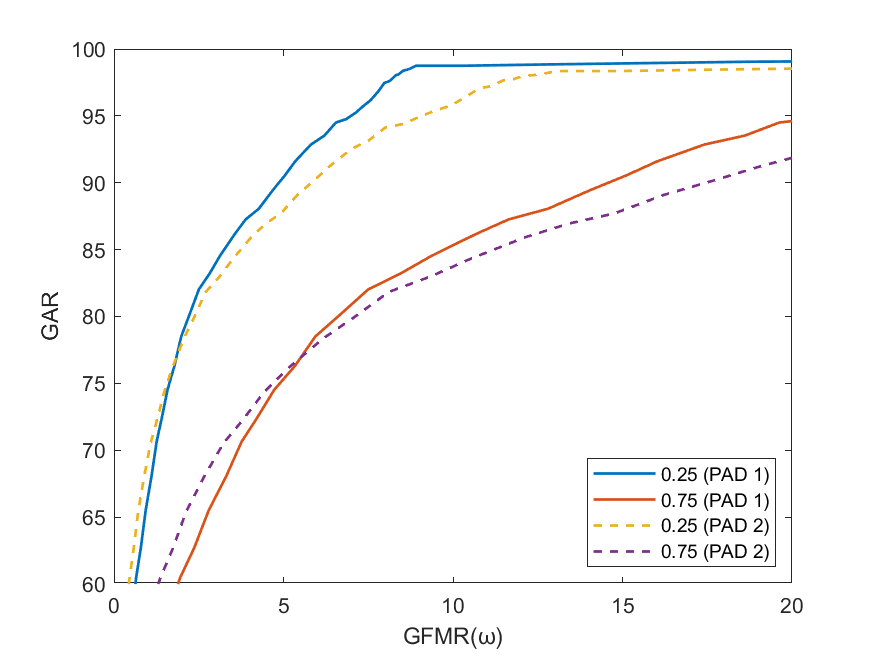}}
\caption{DigitalPersona dataset from LivDet 2019. GROC comparison between the top two LivDet 2019 winners, embedded with Bozorth3 (a,b) and Verifinger 12 (c,d) matching system, when the presentation attacks probability $w \in \{0.25,0.75\}$. Both operational points are reported for each matcher.}
\label{fig:DP_padvspad}
\vspace{-10pt}
\end{figure*}

We tested both algorithms on the most problematic sensor in the competition, namely the DigitalPersona, although our findings can be extended to all the analyzed datasets.
Similar to the previous experiments, we report the results obtained by varying both the PAD operational point ($BPCER_{01}$ or $APCER_{01}$) and the matcher (Bozorth3 or Verifinger 12).
Figure \ref{fig:DP_padvspad} shows the outcome of the comparison. For each plot, we show two attack scenarios: one with a moderately low risk ($w = 0.25$), and one with an intense threat level ($w = 0.75$). 
As can be seen, they confirm the thesis expressed up to now. The degradation in the overall system's performance is due to the different types of PAD and is independent of the matcher since the variations between graphs \ref{fig:DP_padvspad}\{a, c\} and \ref{fig:DP_padvspad}\{b, d\} are the same. This effect is all the more marked, the more stringent the liveness threshold is.
Consequently, thanks to the simulator, it is possible to verify which is the most suitable solution in terms of accuracy for a specific application that employs the sensor under consideration.\newline

\begin{figure}
   \centering
   \subfigure[]{\includegraphics[width=.32\textwidth]{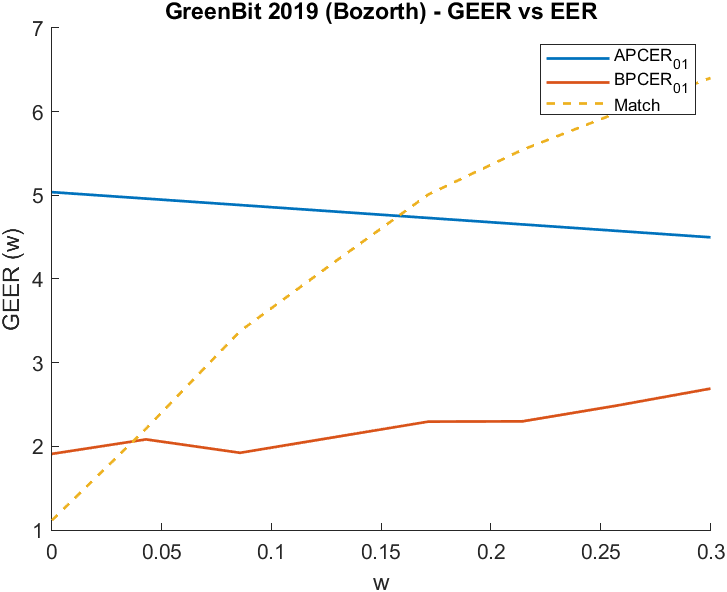}}
   \centering
   \subfigure[]{\includegraphics[width=.32\textwidth]{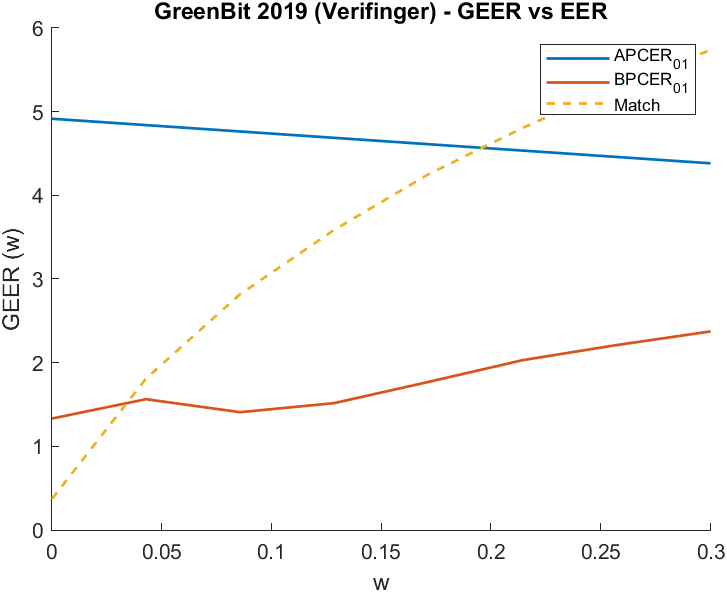} \label{fig:GEERcrv_b}}
      \subfigure[]{\includegraphics[width=.32\textwidth]{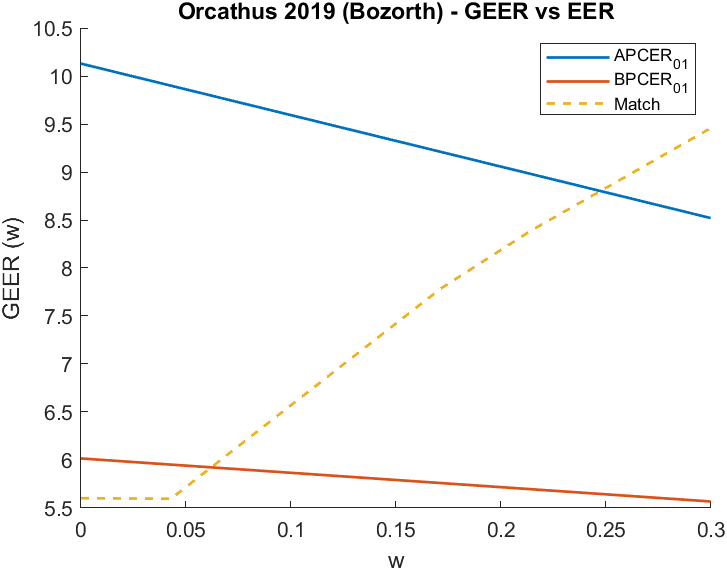}} \\
   \subfigure[]{\includegraphics[width=.32\textwidth]{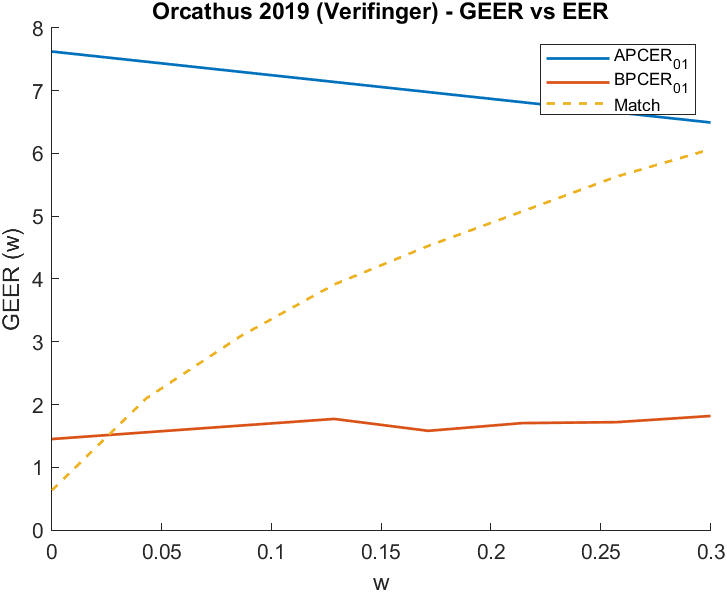}}
      \subfigure[]{\includegraphics[width=.32\textwidth]{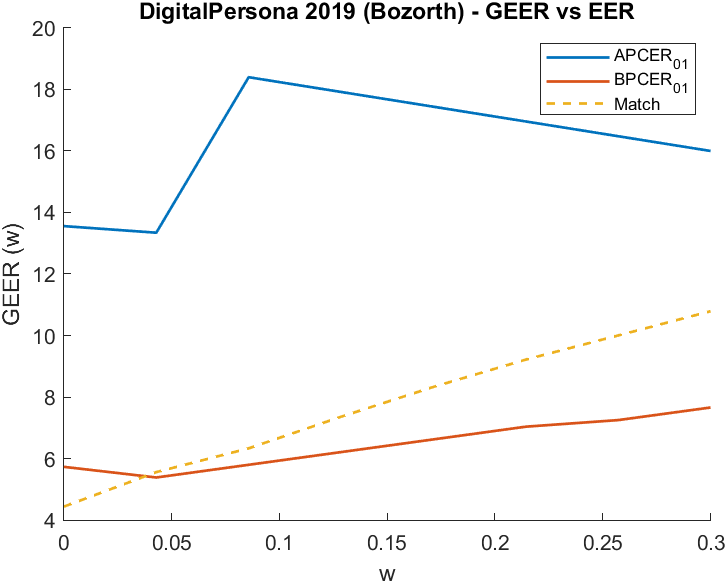}}
   \centering
   \subfigure[]{\includegraphics[width=.32\textwidth]{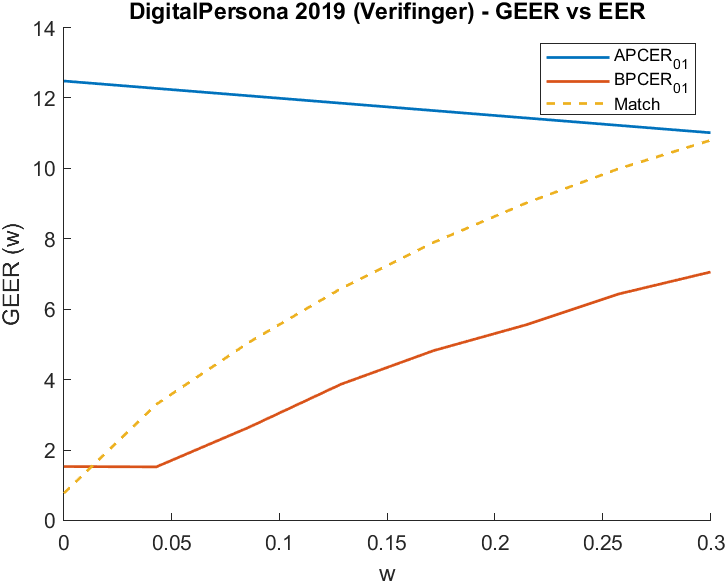}}
\caption{LivDet 2019 datasets: GEER trend when the presentation attacks probability $w \in [0.00,0.30]$. Both matchers are reported: Bozorth3 (left column) and Verifinger 12 (right column). The intersection point of the integrated system curves (solid orange and blue) with the individual system one (dashed yellow) is called $GEER^{*}$. The employed PAD is the "PADUnkFv" algorithm.}

\label{fig:GEERcrv}
\end{figure}

\subsubsection{Application tips: GEER analysis}
\label{subsec:PADdep}

The analyses performed so far aim to help the designer compare several detectors and matchers to choose the most suitable ones according to the final operating context. 
We present in the following an additional, well-established criterion to give a more practical instrument in deciding for which operational points the given embedding is worthy of being implemented or not: the Equal Error Rate (EER). As a matter of fact, the introduction of the GFMR metric allows us to consider the integrated system as a binary classification problem, and therefore we can define the so-called Global-EER as follows:
\begin{eqnarray}
\label{eq:finalModel2}
GEER(\tau ^{*}) = \frac{(GFMR(\tau ^{*})+ (1-GAR(\tau ^{*})))}{2}
\end{eqnarray}
where $\tau ^{*}$ is the optimal threshold which ensures the minor difference between the GFMR and the complementary of the GAR.
Let us now consider the GEER of the integrated system working at $APCER_{01}$ and $BPCER_{01}$ operational points and the EER of the verification system alone, varying the probability of attack $w$. 
We report in Figure \ref{fig:GEERcrv} the corresponding graphs for all LivDet 2019 datasets and matchers examined. 
Besides confirming the observation previously presented in terms of the behavior of the integrated and individual system, this representation allows us to identify precisely a possible intersection point per operational point, called $GEER^{*}$, whose value can be exploited as a parameter for choosing the best solution between the two approaches.
In particular, given ${w}^{*}$ the value of $w$ corresponding to the $GEER^{*}$ point, and $\widehat{w}$ the probability of attack estimated \textit{a priori} for a specific application, if the following condition occurs:
\begin{eqnarray}
\label{eq:finalModel2}
w ^{*} <= \widehat{w}
\end{eqnarray}
then the designer's choice will fall on the integrated system. In other words, the $GEER^{*}$  index is the point at which the PAD begins to improve the performance of the integrated system.
Accordingly, the designer now has a concrete and appropriate instrument to investigate the feasibility of such integration quantitatively.
For example, if we consider the GreenBit sensor, with Verifinger 12 equipped and the PAD working at the $APCER_{01}$ (Fig. \ref{fig:GEERcrv_b}, the ${w}^{*}$ value corresponding to the intersection point of the GEER and EER curves is approximately equal to 0.20. Therefore, if the estimated attack probability should prove to be lower, the PAD can be turned off/replaced or its operational point changed.

\section{Discussions}


\subsection{Contributions and limitations of this work}
Previous Sections showed that the proposed model was able to predict the performance of the sequential combination of a presentation attacks detector and a fingerprint verification system. 
Although limited to the sequential fusion of the above modules, this contribution was not yet present in the literature. Worth remarking, the sequential fusion is one among several possibilities, but it is also the simplest and most frequently adopted. Moreover, it is also the most flexible since one module can be replaced or updated without impacting the other.

By our simulator, the prediction of the error rates can be made \textit{a priori}, that is, before implementing the system on overall. This property can be successfully used in the meta-design process. Given two possible ROC curves (one for the matching system and one for the presentation attacks detector), our model can simulate the embedded performance by overcoming the problem of exact estimation of the individual ROC curves. 

Besides the contributions above, one may object that the proposed simulator works only for the sequential combination of PAD and verification system. To confirm this, we checked the consistency of the simulator in predicting the performance of integrated system algorithms proposed in LivDet2019, which were not based on sequential fusion. 
Unfortunately, the obtained estimation error did not allow a reliable adoption of the model.  As it was largely expected, this simulator has no general application.

However, thanks to the reported experimental analysis, we were able to derive the following guidelines for the designer: 

\begin{itemize}
\item The $w$ parameter gives a precise picture of the integrated system's performance from the point of view of the probability of a spoofing attack compared to zero-effort attacks. Accordingly, the designer can set several values of $w$ to assess different security scenarios. Then, she/he can check whether the system's possible performance is around an acceptable range set by design constraints. If not, the PAD can be replaced or turned off.
\item By selecting different operational points of the liveness detector ($p$ in Fig. \ref{fig:simulator}), it is possible to observe how the system's acceptance rates ($FMR$, $GAR$, and $IAPMR$) vary; moreover, it can be observed if they are within a certain tolerance range, also set by design constraints. Out of them, the PAD can be replaced or turned off.
\item The simulator, jointly with the PAD's ROC associated with different materials or sensors, allows to assess worst-case and best-case scenarios and set the most suitable PAD accordingly. In particular, pay attention to gelatin-like materials and small surface-based sensors (Sections \ref{subsec:sensordep} -- \ref{subsec:materialdep}).
\item A better PAD can be of much more impact than a better matcher. As shown in Section \ref{subsec:PADdep}, the PAD is the most relevant cause of acceptance errors. 
Although experiments explicitly devoted to confirming such hypotheses are necessary, we can observe based on the scientific and technological SOTA: 1) investigated matchers are all minutiae-based. This technology is for sure the most reliable and mature. However, we cannot say wheather different results may be achieved using different matchers based on other features (textural, filters). Early works showed that filter-based algorithms for fingerprint matching are not robust as minutiae-based \cite{prabhakar2002}. Extending this conclusion to deep-learning-based matchers is complex and out of the scope of this paper; 2) PAD systems are based mainly on the training-by-example approach, the accuracy of which depends on several factors as the training set representativeness, avoiding overfitting, etc. Moreover, since in PAD the input pattern may exhibit significant differences concerning those adopted for training (never-seen-before attacks), the response of these systems may be very unexpected and related performances less robust than matchers, where representativeness is strictly defined in terms of the unicity of the subject’s fingerprint.

Therefore, the designer must focus on different PAD solutions and choose a flexible way of integration, for example, allowing to turn off the PAD or relaxing its liveness threshold in case of increasing error rates observed during the system's operations.
\end{itemize}

\subsection{Concluding remarks}
\label{concl}

This paper's central question, that is, \textit{are we ready to embed fingerprint PAD into verification systems?}, was raised by the fact that this embedding showed an overall error rate more significant than that claimed by vendors of fingerprint verification algorithms without such feature, in the recent past.
The proposed simulator showed that answering this question depends on the final operating context. Therefore, we cannot answer in an absolute sense. The reported results suggest that integrating a PAD into a fingerprint verification system is suitable if the operating point is chosen carefully and the probability of an attack is small but non-zero. This is particularly evident in the GEER analysis shown in Figure \ref{fig:GEERcrv_b}. 
On the other hand, the PAD technology seems not yet robust enough to certain materials and sensors, especially when the attack probability couples with the PAD's operational point moving from low BPCER to low APCER. 

In our opinion, PAD technology is mandatory in high-security applications. In this case, both APCER and GFMR must be as lowest as possible, under a high probability of being attacked. However, our results showed that an overall and, often, significant performance loss must be taken into account and evaluated as acceptable or not by the designer. Overall, the PAD technology does not yet appear as fully mature in this context.

In consumer applications, adding a PAD module is suggested when rare and yet very effective attacks lead to significant troubles, such as hacking personal bank accounts or something of the kind. Since the BPCER value can be set to low values due to the rarity of the adverse event, the overall system can benefit from being in the ROC's area where the PAD's embedding significantly improves the performance. 

\section*{References}
\bibliographystyle{elsarticle-num}



\end{document}